\documentclass[a4paper, 12pt, twoside]{Etuthesis}

\usepackage[square, numbers, comma, sort&compress]{natbib}  
\usepackage[turkish]{babel}
\usepackage[utf8]{inputenc}
\usepackage[T1]{fontenc}
\usepackage{verbatim}  
\usepackage{vector}  
\usepackage{mathptmx}
\usepackage{fancyhdr}
\usepackage{tocloft}
\usepackage{titlesec}
\usepackage[format=hang]{caption}
\usepackage{graphicx}
\usepackage{chngcntr}
\usepackage{subcaption}
\usepackage{fixltx2e}

\usepackage{array}
\usepackage{tabularx}
\usepackage{floatrow}
\usepackage{newfloat}
\usepackage{lscape}
\usepackage[pass]{geometry}
\usepackage{multibib}
\usepackage{SIunits}
\usepackage{url}
\usepackage{xspace}
\usepackage{afterpage}

\let\oldcleardoublepage\cleardoublepage
\usepackage{emptypage}

\let\oldsection\section
\def\section{\cleardoublepage\oldsection}

\makeatletter
\renewcommand\NAT@bibsetnum[1]{\settowidth\labelwidth{\@biblabel{#1}}%
    \setlength{\leftmargin}{\bibindent}\addtolength{\leftmargin}{\dimexpr\labelwidth+\labelsep\relax}%
    \setlength{\itemindent}{-\bibindent}%
    \setlength{\listparindent}{\itemindent}
 \setlength{\itemsep}{\bibsep}\setlength{\parsep}{\z@}%
    \ifNAT@openbib
        \addtolength{\leftmargin}{\bibindent}%
        \setlength{\itemindent}{-\bibindent}%
        \setlength{\listparindent}{\itemindent}%
        \setlength{\parsep}{0pt}%
     \fi
     }
\makeatother

\usepackage{tikz}
\newcommand*\circled[1]{\tikz[baseline=(char.base)]{
            \noindent \node[shape=circle,draw,fill=black,text=white,inner
    sep=0.5pt] (char) {#1};}}

\usepackage{pgfplots}
\usepackage{pgfplotstable}
\pgfplotsset{width=10cm, compat=1.5}

\usepgfplotslibrary{external}
\tikzexternaldisable

\titlespacing\subsection{0pt}{12pt plus 4pt minus 2pt}{2pt plus 2pt minus 2pt}

\renewcommand\section{\if@openright\cleardoublepage\else\clearpage\fi
    \thispagestyle{plain}
        \vspace*{1.0cm}

        \oldsection
    }

\usepackage[none]{hyphenat}

\newcommand{\cmdread}{\texttt{{READ}}\xspace}
\newcommand{\cmdwrite}{\texttt{{WRITE}}\xspace}

\newcommand{\cmdras}{\texttt{{RAS}}\xspace}
\newcommand{\cmdcas}{\texttt{{CAS}}\xspace}
\newcommand{\cmdwe}{\texttt{{WE}}\xspace}
\newcommand{\cmdcke}{\texttt{{CKE}}\xspace}
\newcommand{\cmdcs}{\texttt{{CS}}\xspace}

\addto\captionsturkish{\renewcommand{\figurename}{Figure }}
\addto\captionsturkish{\renewcommand{\bibname}{REFERENCES}}

\graphicspath{{figures/}}

\titleformat{\chapter}[display]
  {\normalfont\bfseries}{}{0pt}{}
\titleformat{\section}
  {\normalfont\bfseries}{\textbf{\arabic{section}. }}{1pt}{} 
\titleformat{\subsection}
  {\normalfont\bfseries}{\textbf{\arabic{section}.\arabic{subsection} }}{1pt}{} 
\titleformat{\subsubsection}
  {\normalfont\bfseries}{\textbf{\arabic{section}.\arabic{subsection}.\arabic{subsubsection} }}{1pt}{} 
\titleformat{\paragraph}
  {\normalfont\bfseries}{\textbf{\arabic{section}.\arabic{subsection}.\arabic{subsubsection}.\arabic{paragraph} }}{1pt}{}
\titleformat{\subparagraph}
  {\normalfont\bfseries}{}{1pt}{} 

\renewcommand\thesection{\arabic{section}}
\makeatletter
\renewcommand{\cftsubparapresnum}{\begin{lrbox}{\@tempboxa}}
\renewcommand{\cftsubparaaftersnum}{\end{lrbox}}
\makeatother
\makeatletter
\renewcommand{\numberline}[1]{%
  \@cftbsnum #1\@cftasnum~\@cftasnumb%
}
\cftsetindents{section}{0.5in}{0.5in}
\cftsetindents{subsection}{0.7in}{0.7in}
\cftsetindents{subsubsection}{0.9in}{0.9in}
\cftsetindents{paragraph}{1.1in}{1.1in}
\cftsetindents{subparagraph}{1.3in}{1.3in}

\counterwithin{figure}{section}
\counterwithin{table}{section}
\counterwithin{equation}{section}

\newlength\figurelen
\newlength\tablelen
\AtBeginDocument{%
  \renewcommand\listtablename{\normalsize \vspace{-62pt} \centerline{LIST OF
  TABLES}}
  \renewcommand\listfigurename{\normalsize \vspace{-62pt} \centerline{LIST OF
  FIGURES}}
  \renewcommand\contentsname{\normalsize \hspace{32pt} TABLE OF CONTENTS}
  \renewcommand\tablename{Table}
}

\newcommand{\squishlist} {
        \begin{list}{$\bullet$} {
                \setlength{\itemsep}{-2pt}
                \setlength{\parsep}{2pt}
                \setlength{\topsep}{0pt}
                \setlength{\partopsep}{0pt}
                \setlength{\leftmargin}{1.0em}
                \setlength{\labelwidth}{1em}
                \setlength{\labelsep}{0.5em}
        }
}

\newcommand{\squishend} {
        \end{list}
}

\begin{document}
\thesistitle{DRAM SIZMA KARAKTERİSTİKLERİ VE OLAĞAN ERİŞİM ÖRÜNTÜSÜNDEN
    FAYDALANARAK DRAM ERİŞİM GECİKMESİNİN AZALTILMASI}
\thesistitleeng{REDUCING DRAM ACCESS LATENCY BY EXPLOITING DRAM LEAKAGE
    CHARACTERISTICS AND COMMON ACCESS PATTERNS}	
\thesistype{0}	
\thesisauthor{Hasan HASSAN}
\department{Bilgisayar Mühendisliği Anabilim Dalı}
\departmenteng{Department of Computer Engineering}	
\supervisor{Assoc. Prof. Oguz ERGIN}
\supervisortr{Doç. Dr. Oğuz ERGİN}
\thesisdate{AĞUSTOS 2016}
\thesisdateeng{AUGUST 2016}	
\instno{Enstitü No}

\instdirector{Prof. Osman EROGUL}	
\dephead{Assoc. Prof. Oguz ERGIN}	
\supap{Assoc. Prof. Oguz ERGIN}{TOBB University of Economics and Technology} 
\cosupap{0}{TOBB Ekonomi ve Teknoloji Üniversitesi} 

\juryaptwo{Assoc. Prof. Ali BOZBEY}{TOBB University of Economics and Technology}
\juryapone{Prof. Mehmet Onder EFE (Chair)}{Hacettepe University}

\turabkeys{Devingen Rastgele Erişimli Bellek, Bellek sistemleri.}	
\engabkeys{Dynamic Random Access Memory (DRAM), Memory systems.}	

\thanksprof{Assoc. Prof. Oguz ERGIN}
\thanksdep{Department of Computer Engineering}

\univname{TOBB UNIVERSITY OF ECONOMICS AND TECHNOLOGY}
\univnamesmall{TOBB Ekonomi ve Teknoloji Üniversitesi}
\univnamesmalleng{TOBB University of Economics and Technology}
\instname{INSTITUTE OF NATURAL AND APPLIED SCIENCES}
\instnamesmall{Fen Bilimleri Enstitüsü}
\instnamesmalleng{Institute of Natural and Applied Sciences}
\dummytypemsmall{Yüksek Lisans}
\dummytypembig{MASTERS}
\dummytypedsmall{Doktora}
\dummytypedbig{DOKTORA}
\dummydate{Tarih:}
\dummydateeng{Date:}
\dummythes{THESIS}
\dummysup{Supervisor:}
\dummyinstap{Approval of the Institute of Natural and Applied Sciences}
\dummyinstdir{Director}
\dummyreqapone{I certify that this thesis satisfies all the requirements as a
thesis for the degree of Master of Science.}
\dummyreqaptwo{}
\dummydephead{Deputy Head of Department}

\dummymidapone{
    This thesis entitled \textbf{"REDUCING
            DRAM ACCESS LATENCY BY EXPLOITING DRAM LEAKAGE
                CHARACTERISTICS AND COMMON ACCESS PATTERNS"} has been prepared and
                submitted in partial fulfillment of the requirements for the
                degree of Master of Science in Computer Enginnering by \textbf{Hasan HASSAN}, who is a graduate student at
                TOBB University of Economics and Technology, Institute
                of Natural and Applied Sciences with student number 131111040.
                The thesis has
                been examined in \textbf{AUGUST 11, 2016} by the thesis committee below
                and is recommended for approval and acceptance.
}


\dummymidaptwo{}
\dummymidapthree{}
\dummymidapfour{}
\dummymidapfive{}
\dummysupap{Supervisor:}
\dummysupapeng{Supervisor:}
\dummycosupap{Eş Danışman:}
\dummyjurys{Commitee Members:}
\dummythesacktitle{TEZ BİLDİRİMİ}
\dummythesacktext{Tez içindeki bütün bilgilerin etik davranış ve akademik kurallar çerçevesinde elde edilerek sunulduğunu, alıntı yapılan kaynaklara eksiksiz atıf yapıldığını, referansların tam olarak belirtildiğini ve ayrıca bu tezin TOBB ETÜ Fen Bilimleri Enstitüsü tez yazım kurallarına uygun olarak hazırlandığını bildiririm.}

\dummythesacktitleeng{DECLARATION}
\dummythesacktexteng{I hereby declare that all the information provided in this
thesis has been obtained with rules of ethical and academic conduct and has been
written in accordance with thesis format regulations. I also declare that, as
required by these rules and conduct, I have fully cited and referenced all
material and results that are not original to this work.}

\dummythesacksign{}
\dummyturaboz{ÖZET}
\dummyengaboz{ABSTRACT}
\dummyturabthes{Tezi}
\dummyengabthesm{Master of Science}
\dummyengabthesd{Doctor of Philosophy}
\dummyturabkey{Anahtar Kelimeler:}
\dummyengabkey{Keywords:}
\dummythankstitle{ACKNOWLEDGMENTS}

\dummythanksone{I would like to thank my advisor Oguz Ergin for supporting me in
    every aspect through my undergraduate and graduate education in TOBB
    University of Economics and Technology.  I would not have succeed without
    the priceless knowledge I acquired by working with him. I would also like to
    thank Onur Mutlu and SAFARI for for all the feedback and comments which
    greatly enhanced my research, KASIRGA for creating a stimulating working
    environment, and TOBB University of Economics and Technology for funding me
during my education.  }


\dummythankstwo{}
\dummythanksthree{}

\pagenumbering{roman}

\newgeometry{inner=2.5cm, outer=2.5cm, top=5cm, bottom=2.5cm}
\maketitlex

\newgeometry{inner=4cm, outer=2.5cm, top=5cm, bottom=2.5cm}



\addtocontents{toc}{~\hfill\underline{\textbf{Page}}\par}

\addcontentsline{toc}{section}{\textbf{\tdummyengaboz}}

\begin{engabstract}

    DRAM-based memory is a critical factor that creates a bottleneck on the
    system performance since the processor speed largely outperforms the DRAM
    latency. 

In this thesis, we develop a low-cost mechanism,
called \emph{ChargeCache}, which enables faster access to
recently-accessed rows in DRAM, with no modifications to DRAM
chips. Our mechanism is based on the key observation that a
recently-accessed row has more charge and thus the following
access to the same row can be performed faster. To exploit
this observation, we propose to track the addresses of
recently-accessed rows in a table in the memory controller. If a
later DRAM request hits in that table, the memory controller uses
lower timing parameters, leading to reduced DRAM latency. Row
addresses are removed from the table after a specified duration
to ensure rows that have leaked too much charge are not accessed
with lower latency. We evaluate ChargeCache on a wide variety of
workloads and show that it provides significant performance and
energy benefits for both single-core and multi-core systems.

\end{engabstract}

\cleardoublepage

\addcontentsline{toc}{section}{\textbf{\tdummyturaboz}}
\setstretch{0.90}
\begin{turabstract}

DRAM tabanlı bellek, bilgisayar sisteminde darboğaz oluşturarak sistemin
başarımı sınırlayan en önemli bileşendir. Bunun sebebi işlemcilerin hız
bakımından DRAM'lerin çok önünde olmasıdır. Bu tezde, \emph{ChargeCache} ismini
verdiğimiz, DRAM'lerin erişim gecikmesini azaltan bir yöntem geliştirdik. Bu
yöntem, piyasadaki DRAM yongalarının mimarisinde bir değişiklik
gerektirmediği gibi, bellek denetimcisinde de düşük donanım maliyeti olan
ek birimlere ihtiyaç duymaktadır. ChargeCache, yeni erişilmiş DRAM
satırlarının kısa bir süre sonra tekrar erişileceği gözlemine dayanmaktadır.
Yeni erişilmiş satırlardaki DRAM hücreleri yüksek miktarda yük içerdiğinden,
bunlara hızlı bir şekilde erişilebilir. Bu gözlemden faydalanmak için yeni
erişilen satırların adreslerini bellek denetimcisi içerisinde bir tabloda
tutmayı öneriyoruz. Sonraki erişim isteklerinin bu tablodaki satırlara erişmek
istemesi durumunda, bellek denetimcisi yük miktarı yüksek hücrelerin erişilmek
üzere olduğunu bileceğinden, DRAM erişim değiştirgelerini ayarlayarak erişimin düşük
gecikmeyle tamamlanmasını sağlayabilir. Belirli bir süre sonra tablodaki
satır adresleri silinerek, zaman içerisinde çok fazla yük kaybedip hızlı
erişilebilme özelliğini yitirmiş satırların bu tablodan çıkarılması sağlanır.
Önerdiğimiz yöntemi hem tek çekirdekli hem de çok çekirdekli mimarilerde
benzetim ortamında deneyerek, yöntemin başarım ve enerji kullanımı açısından
sistem üzerinde sağladığı iyileştirmeleri inceledik. 

\end{turabstract}

\cleardoublepage
\setstretch{1.00}

\addcontentsline{toc}{section}{\textbf{\tdummythankstitle}}
\thankspage
\cleardoublepage

\addcontentsline{toc}{section}{\textbf{TABLE OF CONTENTS}}
\tableofcontents
\cleardoublepage

\addcontentsline{toc}{section}{\textbf{LIST OF FIGURES}}
\addtocontents{lof}{~\hfill\underline{\textbf{Page}}\par}
\listoffigures
\vfill
\cleardoublepage

\addcontentsline{toc}{section}{\textbf{LIST OF TABLES}}
\addtocontents{lot}{~\hfill\underline{\textbf{Page}}\par}
\listoftables
\vfill
\cleardoublepage

\addcontentsline{toc}{section}{\textbf{ABBREVIATIONS}}
\section*{\centerline{ABBREVIATIONS}}
\begin{tabular}{ l l }
 \textbf{DRAM} & : Dynamic Random-Access Memory \\ 
 \textbf{SRAM} & : Static Random-access Memory \\ 
 \textbf{DDR} & : Double Data-rate \\ 
 \textbf{LLC} & : Last-level Cache \\ 
 \textbf{RLTL} & : Row-level Temporal Locality \\ 
 \textbf{CMOS} & : Complementary Metal Oxide Semiconductor \\ 
 \textbf{PCB} & : Printed Circuit Board \\ 
 \textbf{I/O} & : Input/Output \\ 
 \textbf{HCRAC} & : Highly-charged Row Address Cache \\ 
\end{tabular}
\vfill
\cleardoublepage



\pagenumbering{arabic}
{
\setlength{\parskip}{10pt}

\renewcommand\cleardoublepage\oldcleardoublepage

\newgeometry{inner=4cm, outer=2.5cm, top=2.5cm, bottom=2.4cm}



\section{INTRODUCTION}
\label{section:intro}

In the last few decades, new microarchitectural techniques successfully
delivered significant performance improvement to the microprocessors. At the
same time, advances in the manufacturing technology, which shrinked the
transistor size, provided additional processing power mainly by enabling more
transistors to fit to the same die area. On the other hand, capacity of the
memories also increased dramatically but the improvement in the speed of the
memory was not high enough to catch up with the processors. The disparity
between the performance of the processors and memory devices introduced a
system-level bottleneck problem which is typically known as the "memory
wall"~\cite{wulf, wilkes}. In todays multi-core era, that bottleneck is even
exagerrated by the increased bandwidth requirements due to the simultaneously
operating processor cores where each of them generate a significant amount of
memory accesses.

DRAM technology is commonly used as the main memory of modern
computer systems. This is because DRAM is at a more favorable
point in the trade-off spectrum of density (cost-per-bit) and
access latency compared to other technologies like SRAM or flash.
However, commodity DRAM devices are heavily optimized to maximize
cost-per-bit. In fact, the latency of commodity DRAM has not
reduced significantly in the past
decade~\cite{lee,mutluresearch}.

To mitigate the negative effects of long DRAM access latency, existing
systems rely on several major approaches. First, they employ large
on-chip caches to exploit the temporal and spatial locality of memory
accesses. However, cache capacity is limited by chip area. Even caches
as large as tens of megabytes may not be effective for some
applications due to very large working sets and memory access
characteristics that are not amenable to
caching~\cite{palacharla1994evaluating,qureshi2007adaptive,jevdjic2014unison,
  lotfi2012scale, qureshi2007line}. Second, systems employ aggressive
prefetching techniques to preload data from memory before it is
needed~\cite{baer1991effective, srinath2007feedback,
  charney1997prefetching}. However, prefetching is inefficient for
many irregular access patterns and it increases the bandwidth
requirements and interference in the memory
system~\cite{ebrahimi2011prefetch, ebrahimi2009techniques,
  ebrahimi2009coordinated,lee2008prefetch}.  Third, systems employ
multithreading~\cite{thornton1964parallel, smith1978pipelined}.
However, this approach increases contention in the memory
system~\cite{moscibroda2007memory, ebrahimi2011parallel, mutlu08,
  das2013application} and does not aid single-thread
performance~\cite{joao2012bottleneck,
  suleman2009accelerating}. Fourth, systems exploit memory level
parallelism~\cite{mutlu2003runahead, glew1998mlp,
  chou2004microarchitecture, mutlu2005techniques, mutlu08}. The DRAM
architecture provides various levels of parallelism that can be
exploited to simultaneously process multiple memory requests generated
by modern processor
architectures~\cite{tomasulo1967efficient,patt1985hps,mutlu2003runahead,lee2009improving}.
While prior works~\cite{jeong12, mutlu08, ding04, lee2009improving,
  pai1999code} proposed techniques to better utilize the available
parallelism, the benefits of these techniques are limited due to
1)~address dependencies among instructions in the
programs~\cite{avd,clap,liupp}, and 2)~resource conflicts in the
memory subsystem~\cite{kim12,rau1991pseudo}. Unfortunately,
\emph{none} of these four approaches \emph{fundamentally} reduce
memory latency at its source and the DRAM latency continues to be a
performance bottleneck in many systems.

The latency of DRAM is heavily dependent on the design of the
DRAM chip architecture, specifically the length of a wire called
\emph{bitline}. A DRAM chip consists of millions of DRAM cells.
Each cell is composed of a transistor-capacitor pair. To access
data from a cell, DRAM uses a component called \emph{sense
amplifier}. Each cell is connected to a sense amplifier using a
\emph{bitline}. To amortize the large cost of the sense
amplifier, hundreds of DRAM cells are connected to the same
bitline~\cite{lee}. Longer bitlines lead to increase in
resistance and parasitic capacitance on the path between the DRAM
cell and the sense amplifier. As a result, longer bitlines result
in higher DRAM access latency~\cite{lee, lee2015, son}.

One simple approach to reduce DRAM latency is to use shorter
bitlines.  In fact, some specialized DRAM chips~\cite{rldram,
  lldram, sato1998fast} offer lower latency by using shorter
bitlines compared to commodity DRAM chips.  Unfortunately, such
chips come at a significantly higher cost as they reduce the
overall density of the device because they require more sense
amplifiers, which occupy significant area~\cite{lee}. Therefore,
such specialized chips are usually not desirable for systems that
require high memory
capacity~\cite{chatterjee2012leveraging}. Prior works have
proposed several heterogeneous DRAM architectures (e.g., segmented
bitlines~\cite{lee}, asymmetric bank organizations~\cite{son})
that divide DRAM into two regions: one with low latency, and
another with slightly higher latency.  Such schemes propose to map
frequently accessed data to the low-latency region, thereby
achieving lower average memory access latency. However, such
schemes require 1)~non-negligible changes to the cost-sensitive
DRAM design, and 2)~mechanisms to identify, map, and migrate
frequently-accessed data to low-latency regions. As a result, even
though they reduce the latency for some portions of the DRAM chip,
they may be difficult to adopt.

\textbf{Our goal} in this work is to design a mechanism to reduce
the average DRAM access latency without modifying the existing
DRAM chips. We achieve this goal by exploiting two major
observations we make in this thesis.

\textbf{Observation 1.} We find that, due to DRAM bank
conflicts~\cite{kim12,rau1991pseudo}, many applications tend to
access rows that were recently closed (i.e., closed within a very short
time interval). We refer to this form of temporal locality where
certain rows are closed and opened again frequently as
\textit{Row Level Temporal Locality (RLTL)}. 
An important outcome of this observation is that a DRAM row
remains in a \emph{highly-charged} state when accessed for the second
time within a short interval after the prior access. This is
because accessing the DRAM row inherently replenishes
the charge within the DRAM cells (just like a refresh operation
does)~\cite{liu2012raidr,shin,nair2013, chang2014, ghosh2007smart,liu2013experimental}.

\textbf{Observation 2.} The amount of charge in DRAM cells
determines the required latency for a DRAM access. If the amount
of charge in the cell is low, the sense amplifier completes its
operation in longer time. Therefore, DRAM access latency
increases. A DRAM cell loses its charge over time and the charge
is replenished by a refresh operation or an access to the row.
The access latency of a cell whose charge has been replenished
recently can thus be significantly lower than the access latency
of a cell that has less charge.

We propose a new mechanism, called
\textit{ChargeCache}~\cite{hassan2016chargecache}, that reduces
average DRAM access latency by exploiting these two observations. The
\textbf{key idea} is to track the addresses of recently-accessed
(i.e., highly-charged) DRAM rows and serve accesses to such rows with
lower latency.  Based on our observation that workloads typically
exhibit significant Row-Level Temporal Locality (see
Section~\ref{section:motivation}), our experimental results on
multi-programmed applications show that, on average, ChargeCache can
reduce the latency of 67\% of all DRAM row activations.

The operation of ChargeCache is straightforward. The memory controller
maintains a small table that contains the addresses of a set of
recently-accessed DRAM rows. When a row is evicted from the row-buffer, the
address of that row, which contains highly-charged cells due to
its recent access, is inserted into the
table. 

Before accessing a new row, the memory controller checks the table to determine
if the row address is present in the table. If so, the row is accessed with low
latency. Otherwise, the row is accessed with normal latency. As cells leak
charge over time, ChargeCache requires a mechanism to periodically invalidate
entries from the table such that only highly-charged rows remain in it.
Section~\ref{section:charge_cache} describes the implementation of ChargeCache
in detail.

Our evaluations show that ChargeCache significantly improves
performance over commodity DRAM for a variety of workloads. For 8-core
workloads, ChargeCache improves average workload performance by 8.6\%
with a hardware cost of only 5.4KB and by 10.6\% with a hardware cost
of 43KB. As ChargeCache can only \emph{reduce} the latency of certain
accesses, it does {\em not} degrade performance compared to commodity
DRAM.  Moreover, ChargeCache can be combined with other DRAM
architectures that offer low latency
(e.g.,~\cite{kim12,lee,lee2015,son,choi,seshadri2013,son2014,seshadri2015fast,chang2014})
to provide even higher performance. Our estimates show that the hardware area
overhead of ChargeCache is only 0.24\% of a 4MB cache. Our mechanism
requires no changes to DRAM chips or the DRAM interface.
Section~\ref{section:results} describes our experimental results.

We make the following \textbf{contributions}.

\squishlist

\item We observe that, due to bank
conflicts, many
applications exhibit a form of locality where
recently-closed DRAM rows are accessed frequently. We refer to this as \textit{Row Level
Temporal Locality (RLTL)}(see Section~\ref{section:motivation}).

\item We propose an efficient mechanism, ChargeCache~\cite{hassan2016chargecache}, which exploits
\textit{RLTL} to reduce the average DRAM access latency by requiring changes
\emph{only} to the memory controller. ChargeCache maintains a
table of recently-accessed row addresses and lowers the latency
of the subsequent accesses that hit in this table within a short
time interval (see Section~\ref{section:charge_cache}).

\item We comprehensively evaluate the performance, energy
  efficiency, and area overhead of ChargeCache. Our experiments
  show that ChargeCache significantly improves performance and
  energy efficiency across a wide variety of systems and workloads
  with negligible hardware overhead (see Section~\ref{section:results}).
\squishend

\section{BACKGROUND ON MAIN MEMORY}
\label{section:background}

Memories are fundamental components used in various parts of the computer
systems (e.g., register, cache, buffers, main memory, etc.). A memory system
consists of multiple layers of memory units where each of these units is
optimized to achieve a specific goal to converge to the ideal memory
which utopically has unlimited bandwidth, zero access latency, infinite
capacity, and no cost. Figure~\ref{fig:memory_system} illustrates a typical
memory system that is implemented in modern computer systems. Each memory unit
in scaled to indicate its actual capacity and access latency. In general, low
capacity memory has lower latency compared to a memory unit with higher
capacity. For example, a very limited memory resource, the register file, can
typically be accessed within a single cycle. Whereas, accessing shared caches
may take up to few tens of cycles to complete.

\begin{figure}[ht]
\centering
\includegraphics[width=0.95\linewidth]{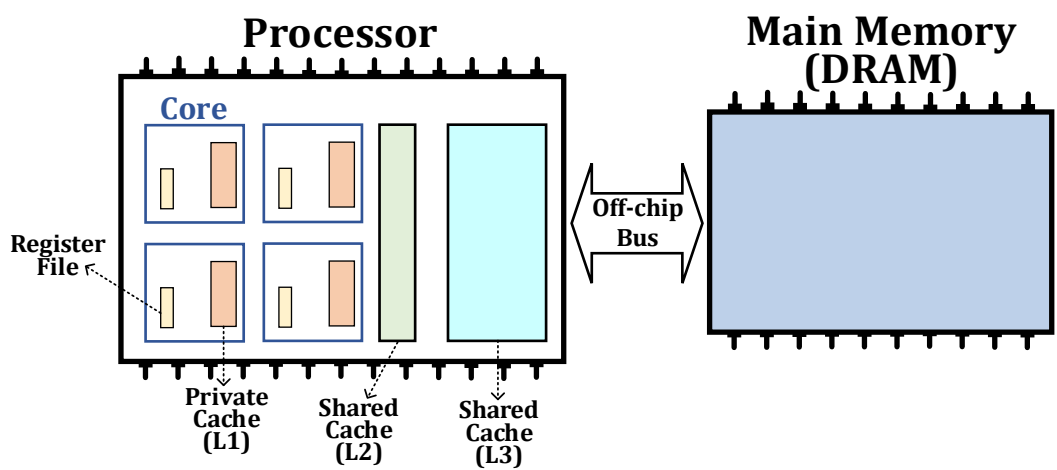}
\caption{Memory system of a modern computer.}
\label{fig:memory_system}
\end{figure}

In this thesis, we mainly focus on the main memory which incurs the highest
access latency in the memory system. DRAM (Dynamic Random Access Memory)
technology is predominantly used as a main memory of modern system. That is
because DRAM is at the most faurable point in the capacity-latency trade-off
spectrum among the memory technologies that are available today. DRAM requires a
special manufacturing process to benefit from its entire potential. Adapting
DRAM to the common CMOS manufacturing technology, which is used to produce the
processor chip (i.e, eDRAM~\cite{matick2005logic}), results in higher area-per-bit
usage and higher access latency compared to a custom-process DRAM. Thus,
in modern systems, DRAM-based main memories are typically available as separate
chip which communicates with the processor via off-chip links. Such a link
imposes additional DRAM access latency.

In this section, we provide the necessary basics on DRAM organization and
operation.


\subsection{DRAM Organization}

DRAM-based main memories are composed of units arranged in hierarchy of several
levels (Figure~\ref{fig:dram_hierarchy}). Next, we explain each level of the hierarchy in detail.

\begin{figure}[!hb] \centering
\includegraphics[width=0.65\linewidth]{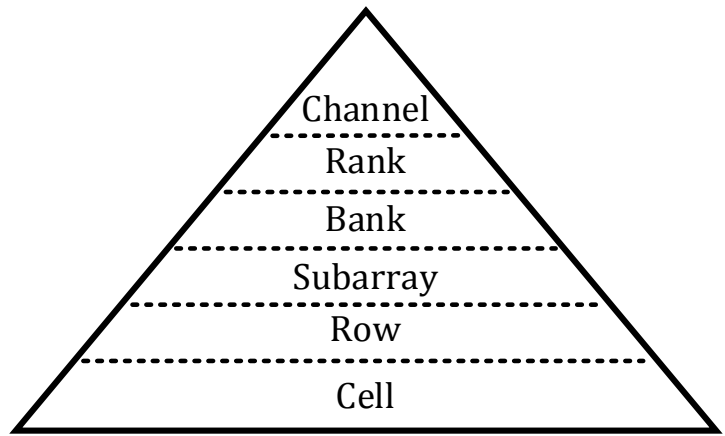}
\caption{Layers of the DRAM hierarchy.}
\label{fig:dram_hierarchy}
\end{figure}

\subsubsection{Channel}

A DRAM channel is the top-level layer of the main memory hierarchy. Each channel
has its own command, address, and data buses. The memory controller, a logic
unit which resides inside the processor chip in modern architectures, handles
the communication with the
channel by issuing a set of DRAM commands to access data in the desired location
(i.e., address). Figure~\ref{fig:dram_channel} shows a system configuration
with two memory controllers which manage a single DRAM channel each. In that
particular system, the workloads running on the processor generate memory
requests. A requests goes to one of the memory controller depending on the
address that it targets. The address space of the system is typically
spread between the two channels. Once a memory controller receives a request, it
issues necessary DRAM commands to the channel to perform the access.

\begin{figure} \centering
\includegraphics[width=0.95\linewidth]{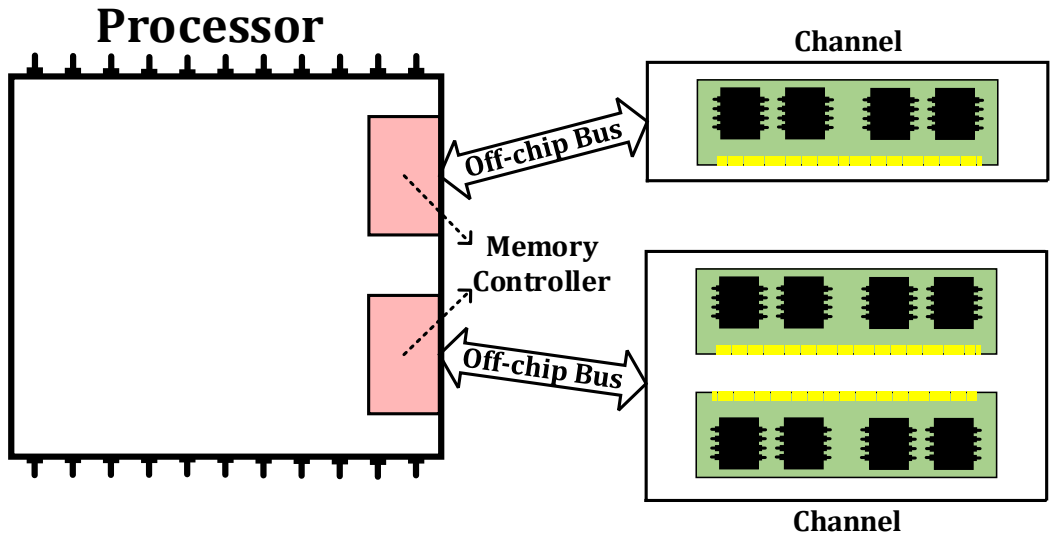}
\caption{View of a system with two channels.}
\label{fig:dram_channel}
\end{figure}

Several DRAM chips are put together to form a DRAM channel. In general-purpose
systems (e.g., desktop computers, laptops, workstations) the chips that create a
channel are solered into a separate PCB (Printed Circuit Board) apart from the
motherboard. These PCBs are called memory modules. A memory module can be
directly plugged to the motherboard through the memory slots. A single channel
may support one or more modules (as in Figure~\ref{fig:dram_channel}). If more
than one modules are connected to a single channel, each module operates as a
DRAM Rank which we explain next. Said that, a channel may contain one or more
ranks (typically up to 4 ranks). On the other hand, in embedded systems (e.g.,
smartphones, single-board computers), DRAM chips are generally soldered to the
motherboard along with other chips of the system.  

\subsubsection{Rank}
Different from channels, ranks do not operate in complete isolation from each
other. Ranks that constitute the same channel share the address, data, and
command buses (Figure~\ref{fig:dram_rank}). Therefore, the ranks operate in
lock-step (i.e., the ranks of the same channel are time multiplexed) and do not offer pure memory
access parallelism as the channels do. However, the ranks offer parallelism in
lower levels of the DRAM hierarchy.

\begin{figure} \centering
    \includegraphics[width=0.75\linewidth]{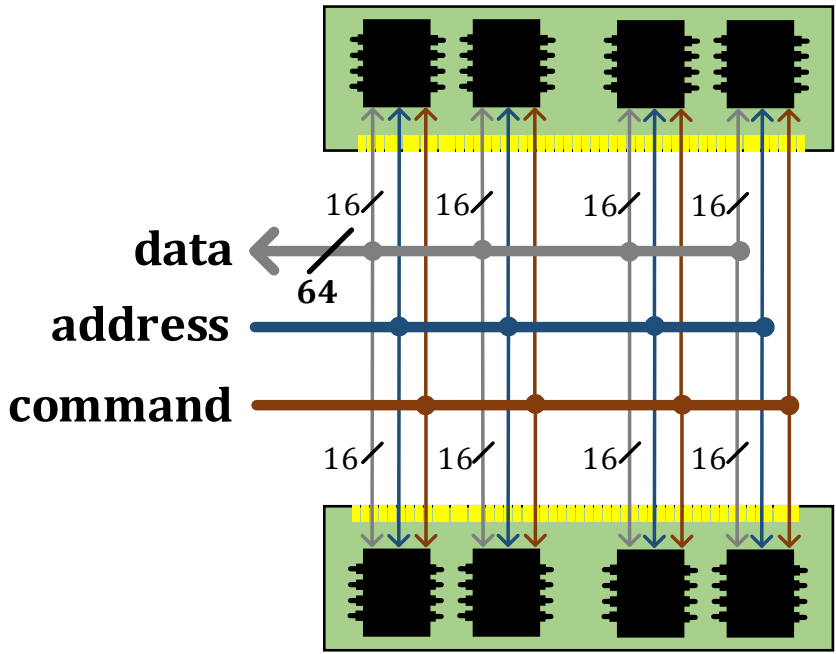}
\caption{A channel which has two ranks that share data, command, and address
buses.}
\label{fig:dram_rank}
\end{figure}

Ranks are composed of multiple DRAM chips. The number of chips depend on the
data I/O width of the used chips and the width of the memory controller bus. In
typicaly systems, the memory controller data bus is 64-bits wide. To reach the
data bus width, multiple chips operate concurrently in a rank. For example, 4
DRAM chips with 16 data I/O pins each are required to form a rank.  

\subsubsection{Bank}

In each rank, there are typically 8 banks available which mostly operate
independently of each other. As shown in Figure~\ref{fig:dram_bank} banks share the same I/O
interface. They utilize that interface in lock-step fashion.
The memory controller, which is on the other side of the I/O bus, can read/write
to/from only one bank at once. Similarly, a data access command mostly targets
a single bank. Some commands (used to initiate operations such as refresh and
precharge) may apply to the all banks in a rank.

\begin{figure}[!hb] \centering
\includegraphics[width=0.65\linewidth]{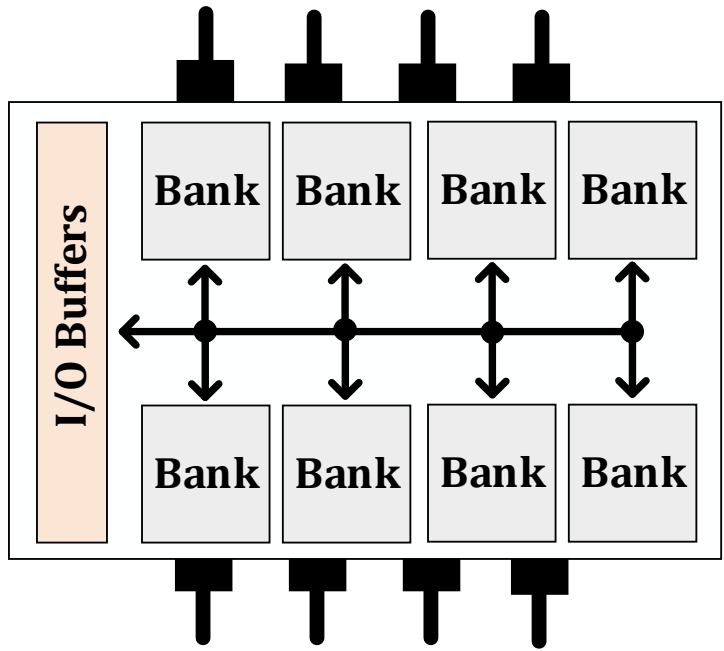}
\caption{The internal structure of a rank which has 8 banks.}
\label{fig:dram_bank}
\end{figure}

Each memory cycle, only a single bank can receive a data access command.
However, since the access operation takes more than one cycle, issuing access
commands to different banks consecutively enables utilization of multiple banks.
For example, assume that an access takes 10 cycles to complete. After issuing an
access command to the first bank, in the next cycle the memory controller may
issue command to serve a request whose data is in different bank. This way, the
latency of two accesses can be overlapped. Overlapping the access time of
multiple requests that go to different banks is called \emph{Bank-Level
Parallelism}. It is critical to exploit the bank-level paralelism to achieve
high throughput~\cite{ding04, lee2009improving, mutlu08, jeong12, kim2010thread,
kim2010}.

\subsubsection{Subarray and row}

Figure~\ref{fig:dram_row} depics a DRAM bank. A bank is composed of several
subarrays and a global row-buffer. Each subarray has hundreds of DRAM rows and a
local row-buffer. Rows are connected to the local row-buffers via local
bitlines. Similarly, local row-buffers are wired to the global row-buffer via
global bitlines. The rows in a bank are grouped into subarrays to keep bitlines
shorter and improve access latency by mitigating parasitic bitline capacitance.
Subarrays do not provide any parallelism in current commercially available
architectures. However, recent work proposes an efficient way to enable
additional level of DRAM parallelism by exploiting subarray
structure~\cite{kim12}.

\begin{figure} \centering
\includegraphics[width=0.85\linewidth]{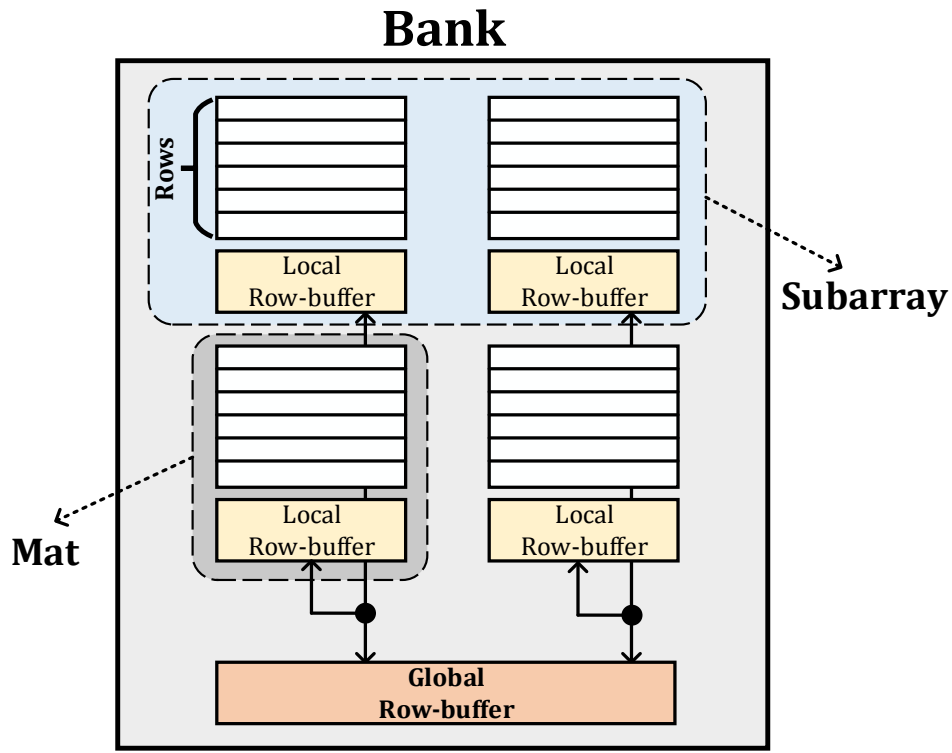}
\caption{The internal structure of a bank.}
\label{fig:dram_row}
\end{figure}

To perform a data access, the row that
corresponds to the accessed address must be first opened by copying that row to
the local row-buffer. After the data is put to the local row-buffer, the data is
transferred to the global row-buffer. Opening a row is also called
\emph{Activation}. Once the data arrives the global row-buffer, the memory
controller can fetch or modify a needed chunk, called column, of the global
row-buffer using a single read or write command. The width of the column depends
on the data I/O width of the DRAM chip.

\subsubsection{Cell}
A DRAM cell consists of a single transistor-capacitor pair. The capacitor stores
a single bit of data depending on its charge level. Asserting the wordline
enables the transistor (i.e., access transistor) which couples up the capacitor
and bitline. Such an operation is necessary to access a DRAM cell.

Due to the One-Transistor One-Capacitor (1T1C) architecture, a DRAM cell faces a
critical leakage problem. Both the transistor and capacitor continuously leak
significant amount of current which causes the DRAM cell to lose its data in
milisecond-long time. As a workaround, the memory controller periodically
initiates a refresh operation which restores the charge level of the cells.

\subsection{DRAM Standards}

Joint Electron Device Engineering Council (JEDEC)~\cite{jedec} defines standards
for manufacturing a wide-range of electronic devides. JEDEC standards also
involve DRAM-based memories. For example, Double Data Rate
(DDR)~\cite{micronDDR3} and its derivatives (such as DDR2, DDR3, DDR4) are the
most widely adopted standards in DRAM memory devices. Other standards such as
High Bandwidth Memory (HBM)~\cite{hbm2013}, Wide I/O DRAM~\cite{dutoit20130},
Low-power DDR (LPDDR)~\cite{jesd2092007low}, and Reduced Latency DRAM
(RLDRAM)~\cite{rldram} are also available. As an example for a DRAM standard, we
briefly explain the DDR3 specification which we use to evaluate our mechanism.

\subsubsection{Double data rate type 3 (DDR3)}

DDR3 standard defines a pin-interface which supports a set of commands that the
memory controller uses to access (e.g., \textit{ACT}, \textit{PRE},
\textit{READ}, \textit{WRITE}) and manage (e.g., \textit{REF}) the memory in a
way we explain in Section~\ref{subsection:dram_op}.

DDR commands are transmitted to the DRAM module across the memory command bus. Each
command is encoded using five output signals (\cmdcke, \cmdcs, \cmdras, \cmdcas,
and \cmdwe). Enabling/disabling these signals corresponds to specific commands
(as defined by the DDR standard). First, the \cmdcke signal ({\em clock
enable}) determines whether the DRAM is in ``standby mode'' (ready to be
accessed) or ``power-down mode''. Second, the \cmdcs ({\em chip selection})
signal specifies the chip that should receive the issued command.  Third, the
\cmdras ({\em row address strobe})/\cmdcas ({\em column address strobe}) signal
is used to generating commands related to DRAM row/column operations.
Fourth, the \cmdwe signal ({\em write enable}) in combination with \cmdras and
\cmdcas, generates the specific row/column command. For example, enabling
\cmdcas and \cmdwe together generates a \cmdwrite command, while only enabling
\cmdcas indicates a \cmdread command.

\subsection{DDR3 Operation}
\label{subsection:dram_op}

\afterpage{
    \clearpage
    \thispagestyle{empty}
    \begin{landscape}
        \centering
        \vspace*{\fill}
        \begin{figure}[htpb]
            \centering
            \includegraphics[width=\linewidth]{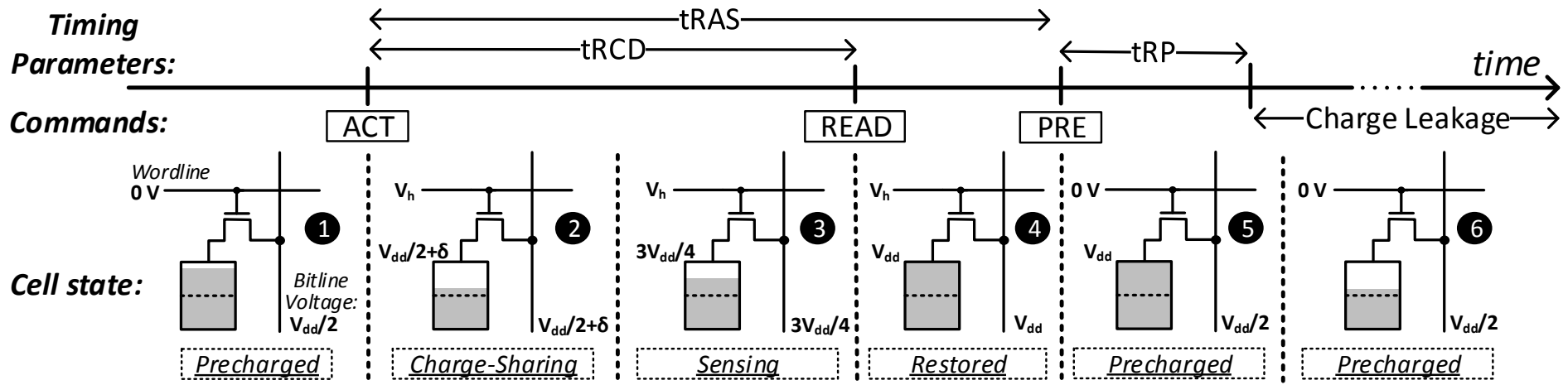}
            \caption{Commands that are used to read data from DRAM and the timing parameters
                associated with them}
            \label{figure:cmd_timeline}
        \end{figure}
        \null\vfill\centerline{\raisebox{-2cm}{\thepage}}
    \end{landscape}
    \clearpage
}

DDR3 provides a set of commands which are used to perform a read/write access or
other operations such as refresh. The memory controller issues these commands
in specific order with certain amount of delay in between to complete the
intended operation. The timing delay that must be respected between certain
command is referred to as \emph{DRAM Timing Parameters}. We explain the commands
and timing parameters used to perform a typical read/write operation.

Figure~\ref{figure:cmd_timeline} shows the different sub-steps involved in
transferring the data from a DRAM cell to the sense amplifier and their mapping
to DRAM commands. Each sub-step takes some time, thereby imposing some
constraints (i.e., timing parameters) on when the memory controller can issue
different commands. The figure also shows the major timing parameters that
govern regular DRAM operation.

In the initial \emph{precharged} state \circled{1}, the bitline is precharged to
a voltage level of V\textsubscript{dd}/2. The wordline is lowered (i.e., at 0V)
and hence, the bitline is \emph{not} connected to the capacitor. An access to
the cell is triggered by the \textit{ACT} command to the corresponding row. This
command first raises the wordline (to voltage level V\textsubscript{h}), thereby
connecting the capacitor to the bitline. Since the capacitor (in this example)
is at a higher voltage level than the bitline, charge flows from the capacitor
to the bitline, thereby raising the voltage level on the bitline to
V\textsubscript{dd}/2$+\delta$ ~\circled{2}. This phase is called \emph{charge
sharing}. After the charge sharing phase, the sense amplifier is enabled and it
detects the deviation on the bitline, and amplifies the deviation. This process,
known as \emph{sense amplification}, drives the bitline and the cell to the
voltage level corresponding to the original state of the cell
(V\textsubscript{dd} in this example). Once the sense amplification has
sufficiently progressed \circled{3}, the memory controller can issue a
\textit{READ} or \textit{WRITE} command to access the data from the cell. The
time taken by the cell to reach this state \circled{3} after the \textit{ACT}
command is specified by the timing constraint \textit{tRCD}. Once the sense
amplification process is complete \circled{4}, the bitline and the cell are both
at a voltage level of V\textsubscript{dd}. In other words, the original charge
level of the cell is fully-\emph{restored}. The time taken for the cell to reach
this state \circled{4} after the \textit{ACT} is specified by the timing
constraint \textit{tRAS}. In this state, the bitline can be precharged using the
\textit{PRE} command to prepare it for accessing a different row. This process
first lowers the wordline, thereby disconnecting the cell from the bitline. It
then precharges the bitline to a voltage level of
V\textsubscript{dd}/2~\circled{5}. The time taken for the precharge operation is
specified by the timing constraint \textit{tRP}.

\textbf{DRAM Charge Leakage and Refresh.} As DRAM cells are not ideal, they leak
charge after the precharge operation~\cite{liu2012raidr, liu2013experimental}.
This is represented in state \circled{6} of Figure~\ref{figure:cmd_timeline}. As
described in the previous section, an access to a DRAM cell fully restores the
charge on the cell (see states \circled{4} and \circled{5}). However, if a cell
is not accessed for a sufficiently long time, it may lose too much charge that
its last cell state may be flipped. To avoid such cases, DRAM cells are
periodically refreshed by the memory controller using the refresh (\textit{REF})
command. The interval at which DRAM cells should be refreshed by the controller
is referred to as the \emph{refresh interval}.

\subsection{Memory Controller}

The Memory Controller sits betweens the Last-level Cache (LLC) and the DRAM.
Today, the memory controller is typically employed in the same chip with the
processor logic, as show in Figure~\ref{fig:memory_controller}.

\begin{figure}[!h] \centering
\includegraphics[width=0.95\linewidth]{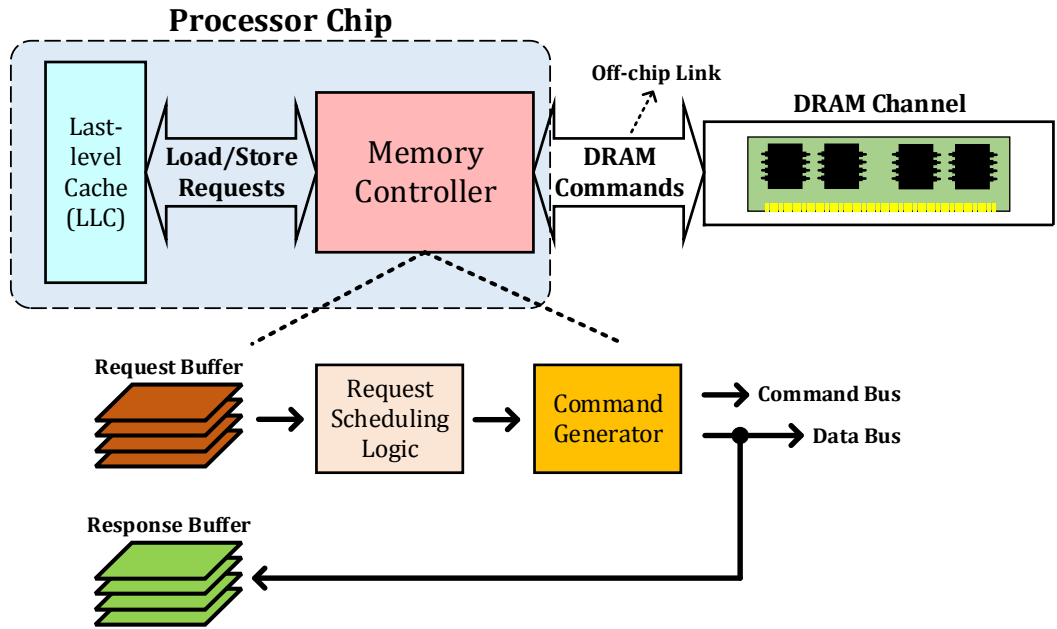}
\caption{Overview of a typical memory controller.}
\label{fig:memory_controller}
\end{figure}

The memory controller is mainly responsible for handling the load/store requests
generated by the LLC. The bottom part of Figure~\ref{fig:memory_controller}
shows an illustration of functional building block of a memory controller. Due
to cache misses or dirty data evictions, LLC generates load/store requests.
Once received, the memory controller stores these requests in the \emph{Request
Buffer}. Then the \emph{scheduling logic} decides which request from the request
buffer to serve first. The scheduling logic (or simply scheduler) makes this
decision based on a set of heuristics which may improve average request serving
time (latency), fairness, or throughput. Once the scheduler makes its decision,
based on the state of the target bank, the \emph{Command Generator} cracks the
request into appropriate DRAM commands. For instance, if the target bank has an
open row and the address of that row is the same as the target row of the
requests, then the command generator only issues a \textit{READ} or
\textit{WRITE} command to the target bank. Whereas, if we have row conflict
(i.e., if the address of the target row is different from the open row address)
the memory controller first issues a \textit{PRE} command to close the
conflicting row. Then, by issuing an \textit{ACT}, the memory controller
activates the target row of the request that is being serviced. Thus, the output
of the command generator not only depends on the decision of the scheduler, but
also on the internal state of the DRAM. The memory controller also receives data
from the DRAM and forwards it to the LLC to respond to the load request.

A memory controller employs smart scheduling algorithms to \emph{(i)} reduce access latency,
\emph{(ii)} improve throuput, or \emph{(iii)} provide better quality of service
(QoS) among concurrently running workloads. A large number of prior work
studiues scheduling algorithms to improve these three aspects~\cite{mutlu08,
jeong12, kim12, frfcfs, nair2013, chang2014, kim2010, kandemir2015, duong2012,
kim2010thread, zhang2000permutation, iyer2007qos, chatterjee2012leveraging,
awasthi2011prediction, ebrahimi2011parallel}. 


\section{MOTIVATION}
\label{section:motivation}

The key takeaway from DRAM operation that we exploit in this work
is the fact that \emph{cells closer to the fully-charged state can
  be accessed with lower activation latency} (i.e., lower
\textit{tRCD} and \textit{tRAS}) than standard DRAM
specification. A recent work~\cite{shin} exploits this observation
to access rows, that were recently recharged via a \emph{refresh}
operation, with lower latency. Specifically, when a row needs to be
activated, the memory controller determines when the row was last
refreshed.  If the row was refreshed recently (e.g., within
$8ms$), the controller uses a lower \textit{tRCD} and
\textit{tRAS} for the activation.

However, this refresh-based approach for lowering latency has two
shortcomings. First, with the standard refresh mechanism, the
refresh schedule used by the memory controller has no correlation
with the memory access characteristics of the application.
Therefore, depending on the point when the program begins
execution, a particular row activation, due to a memory access
initiated by the program, may or may not be to a
recently-refreshed row. Therefore, a mechanism that reduces
latency to recently-refreshed rows cannot provide consistent performance
improvement.
Second, if we use only the time
from the last refresh to identify rows that can be accessed with
low latency (i.e., highly-charged rows), we
find that only 12\% of all memory accesses benefit from low
latency (see Figure~\ref{figure:row_locality}). However, as we
show next, a much greater number of rows can actually be accessed
with low latency.

\afterpage{
    \clearpage
    \thispagestyle{empty}
    \begin{landscape}
        \centering
        \vspace*{\fill}
        \begin{figure}[htpb]
            \centering
            \begin{subfigure}{\linewidth}
                \centering
                \vspace{-15mm}
                \includegraphics[width=\linewidth]{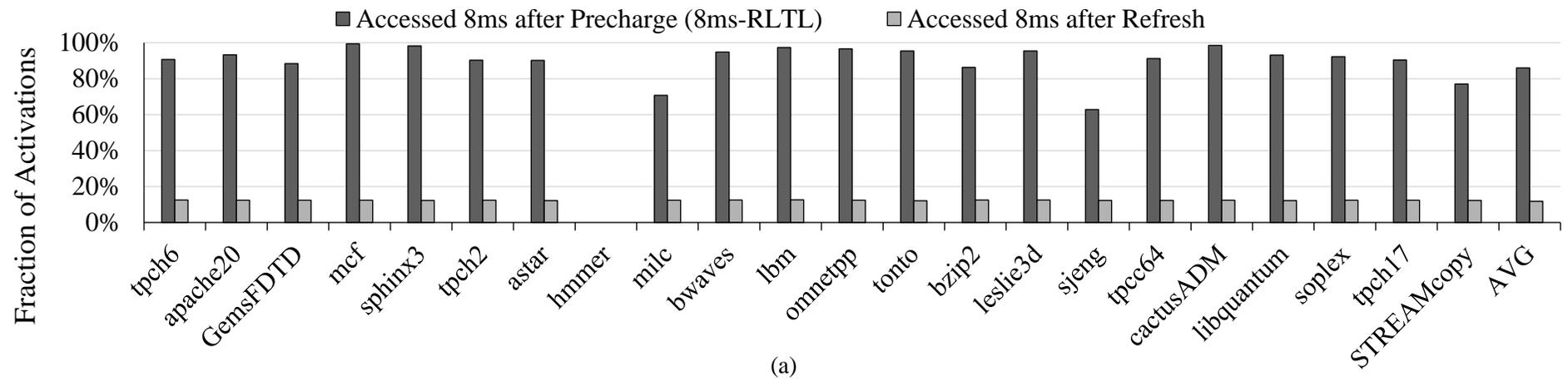}
                \vspace{-10mm}
                \caption{}
                \label{subfigure:row_locality_a}
                \vspace{5mm}
            \end{subfigure}
            \begin{subfigure}{\linewidth}
                \centering
                \includegraphics[width=\linewidth]{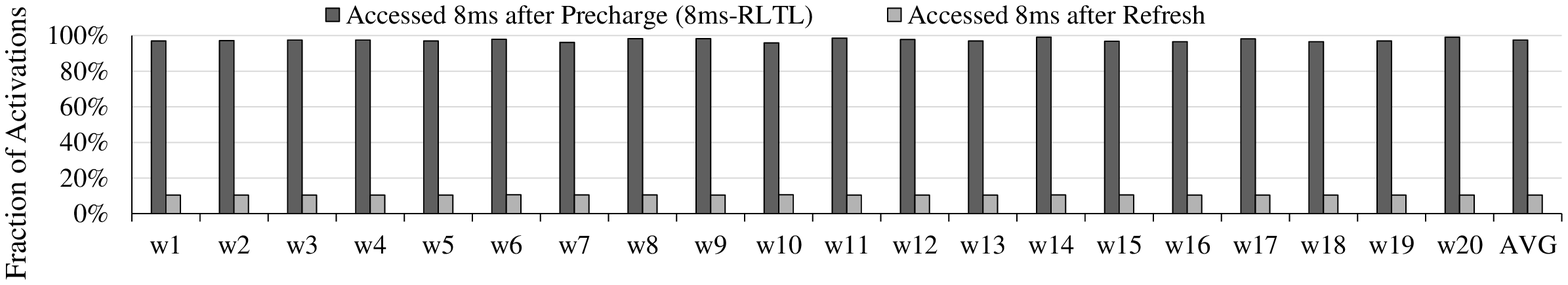}
                \vspace{-8mm}
                \caption{}
                \label{subfigure:row_locality_b}
            \end{subfigure}
    
            \caption{Fraction of row activations that happen 8ms after precharge
                (8ms-RLTL) or refresh of the row ((a) Single-core workloads, (b)
            Eight-core workloads).}
            \label{figure:row_locality}
        \end{figure}
        \null\vfill\centerline{\raisebox{-.4cm}{\thepage}}
    \end{landscape}
    \clearpage
}

As we described in Section~\ref{subsection:dram_op}, an access to
a row fully recovers the charge of its cells. Therefore, if a row
is \emph{activated twice in a short interval}, the second
activate can be served with lower latency as the cells of that
row would still be highly charged. We refer to this notion of row
activation \emph{locality} as \emph{Row-Level Temporal Locality}
(RLTL). We define \emph{t-RLTL} of an application for a given time
interval \emph{t} as the fraction of row activations in which the
activation occurs within the time interval \emph{t} after a previous
precharge to the same row. (Recall that, a row starts leaking
charge only after the precharge operation as shown in
Section~\ref{subsection:dram_op}).

To this end, we would like to understand what fraction of rows exhibit
RLTL, and thus can be accessed with low latency after a precharge
operation to the row due to program behavior versus what fraction of
rows are accessed soon after a refresh to the row and thus can be
accessed with low latency due to a recent preceding refresh.
Figure~\ref{subfigure:row_locality_a} compares the fraction of row
activations of that happen within $8ms$ after the corresponding row is
refreshed to the $8ms$-RLTL of various applications.  As shown in the
figure, with the exception of \emph{hmmer}\footnote{\emph{hmmer}
  effectively uses the on-chip cache hierarchy. Therefore, we do not
  observe any requests to the main memory.}, the $8ms$-RLTL (86\% on
average) is significantly higher than the fraction of row activations
within $8ms$ after the refresh of the row (12\% on average).
Figure~\ref{subfigure:row_locality_b} plots the corresponding values
on an 8-core system that executes 20 multiprogrammed workloads, with
randomly chosen applications for each workload.  As shown, the
fraction of row activations within $8ms$ after refresh is almost the
same as that of the single-core workloads.  This is because the
refresh schedule has no correlation with the application access
pattern. On the other hand, the $8ms$-RLTL for the 8-core workloads is
much higher than that of the single-core workloads. This is because,
in multi-core systems, the exacerbated bank-level
contention~\cite{muralidhara2011reducing, kim2010thread,
  zhang2000permutation,mutlu08,mutlu2007stall, lee2009improving}
results in row conflicts, which in turn results in rows getting closed
and activated within shorter time intervals, leading to a high RLTL.

Figure~\ref{figure:reaccess_hits} shows the RLTL for different
single-core and 8-core workloads with five different time
intervals (from $0.125ms$ to  $32ms$) as a stacked bar and two
different DRAM row management policies, namely, open-row and
closed-row~\cite{awasthi2011prediction,kim2010}.  For each workload, the
first bar represents the results for the open-row policy, and the
second bar represents the results for the closed-row policy. The
open-row policy prioritizes row-buffer hits by keeping the row
open until a request to another row is scheduled (bank
conflict). In contrast, the closed-row policy proactively
closes the active row after servicing all row-hit requests in
the request buffer.

For single-core workloads (Figure~\ref{subfigure:reaccess_a}),
regardless of the row-buffer policy, even the average $0.125ms$-RLTL
is 66\%. In other words, 66\% of all the row activations occur
within $0.125ms$ after the row was previously precharged. For
8-core workloads (Figure~\ref{subfigure:reaccess_b}), due to the
additional bank conflicts, the average $0.125ms$-RLTL is 77\%,
significantly higher than that for the single-core workloads. 
Similar to the single-core workloads, the row-buffer policy does not
have a significant impact on the RLTL for the 8-core workloads.

\afterpage{
    \clearpage
    \thispagestyle{empty}
    \begin{landscape}
        \centering
        \vspace*{\fill}
        \begin{figure}[htpb]
            \centering    
            \begin{subfigure}{.95\linewidth}
                \centering
                \includegraphics[width=\linewidth]{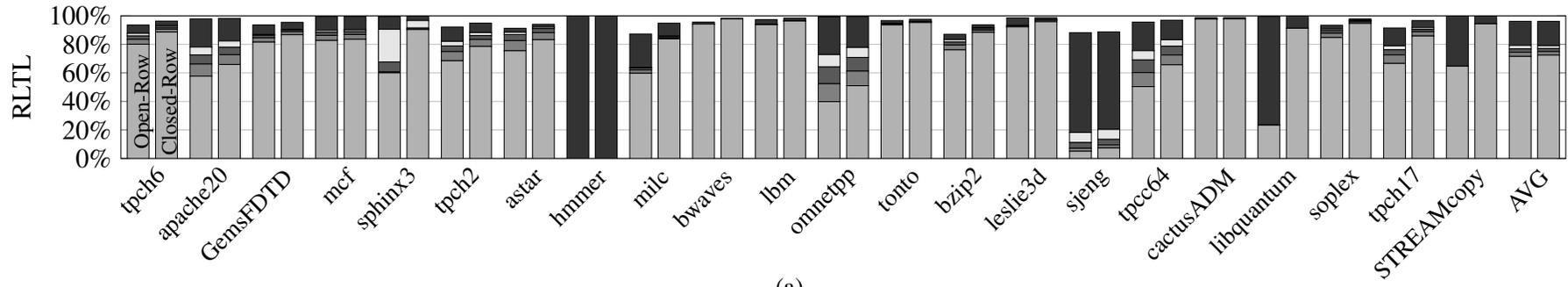}
                \vspace{-12mm}
                \caption{}
                \label{subfigure:reaccess_a}
            \end{subfigure}
            \begin{subfigure}{.95\linewidth}
                \centering
                \includegraphics[width=\linewidth]{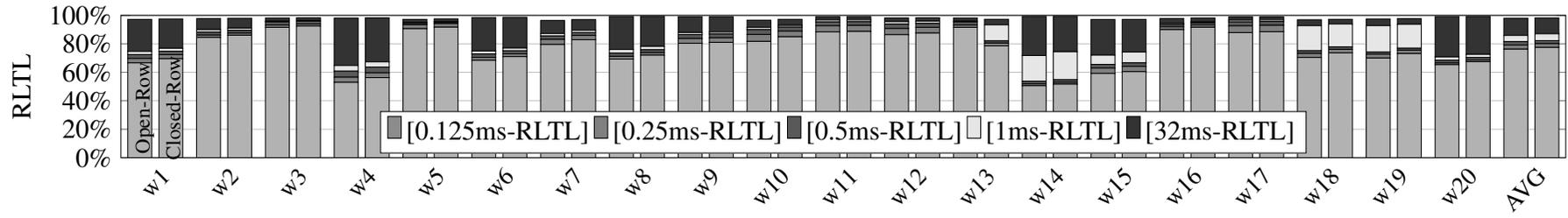}
                \caption{}
                \label{subfigure:reaccess_b}
            \end{subfigure}
            \caption{RLTL for various time intervals ((a) Single-core workloads,
            (b) Eight-core workloads).}
            \label{figure:reaccess_hits}
        \end{figure}
        \null\vfill\centerline{\raisebox{-2cm}{\thepage}}
    \end{landscape}
    \clearpage
}

\textbf{Key Observation and Our Goal.}
We observe that \emph{many} applications exhibit high row-level
temporal locality. In other words, for many applications, a
significant fraction of the row activations occur within a small
interval after the corresponding rows are precharged. As a
result, such row activations can be served with lower activation
latency than specified by the DRAM standard. \textbf{Our goal} in
this work is to exploit this observation to reduce the effective
DRAM access latency by tracking recently-accessed DRAM rows in
the memory controller and reducing the latency for their next
access(es). To this end, we propose an efficient mechanism,
ChargeCache, which we describe in the next
section.

\section{CHARGECACHE}
\label{section:charge_cache}

ChargeCache is based on three observations:
1)~rows that are highly-charged can be accessed with lower
activation latency, 2)~activating a row refreshes the charge on
the cells of that row and the cells start leaking only after the
following precharge command, and 3)~many applications exhibit high
row-level temporal locality, i.e., recently-activated rows are
more likely to be activated again. Based on these observations,
ChargeCache tracks rows that are recently activated, and serves
future activates to such rows with lower latency by lowering the
DRAM timing parameters for such activations.

\subsection{High-level Overview} 
\label{subsection:cc_overview}

At a high level, ChargeCache adds a small table (or cache) to the
memory controller that tracks the addresses of recently-accessed
DRAM rows, i.e., highly-charged rows. ChargeCache performs three
operations. First, when a precharge command is issued to a bank,
ChargeCache inserts the address of the row that was activated in
the corresponding bank to the table
(Section~\ref{subsubsection:hcrac_insertion}). Second, when an
activate command is issued, ChargeCache checks if the
corresponding row address is present in the table. If the address is
not present, then ChargeCache uses the standard DRAM timing
parameters to issue subsequent commands to the bank. However, if
the address of the activated row is present in the table,
ChargeCache employs reduced timing parameters for subsequent
commands to that bank
(Section~\ref{subsubsection:employing_lower_params}). Third,
ChargeCache invalidates entries from the table to ensure that rows
corresponding to valid entries can indeed be accessed with lower
access latency (Section~\ref{subsubsection:invalidating_stale}).

We named the mechanism \emph{ChargeCache} as it provides a
\emph{cache}-like benefit, i.e., latency reduction based on a locality
property (i.e., RLTL), and does so by taking advantage of the
\emph{charge} level stored in a recently-activated row. The
mechanism could potentially be used with current and emerging
DRAM-based memories where the stored charge level leads to different
access latencies. We explain how ChargeCache can be applied to
other DRAM standards in Section~\ref{subsection:other_dram}.

In the following section, we describe the different components and
operation of ChargeCache in more detail. In
Section~\ref{subsection:spice_sim}, we present the results of
our SPICE simulation that analyzes the potential latency reduction that
can be obtained using ChargeCache.

\subsection{Detailed Design} \label{subsection:cc_components}

ChargeCache adds two main components to the memory
controller. Figure~\ref{figure:charge_cache} highlights these
components. The first component is a tag-only cache that stores
the addresses of a subset of highly-charged DRAM rows. We call
this cache the \emph{Highly-Charged Row Address Cache} (HCRAC). We
organize HCRAC as a set-associative structure similar to the
processor caches. The second component is a set of two counters
that ChargeCache uses to invalidate entries from the HCRAC that
can potentially point to rows that are no longer highly-charged.
As described in the previous section, there are three specific
operations with respect to ChargeCache: 1)~insert, 2)~lookup, and
3)~invalidate. We now describe these operations in more detail.

\begin{figure}
\centering
\includegraphics[width=.95\linewidth]{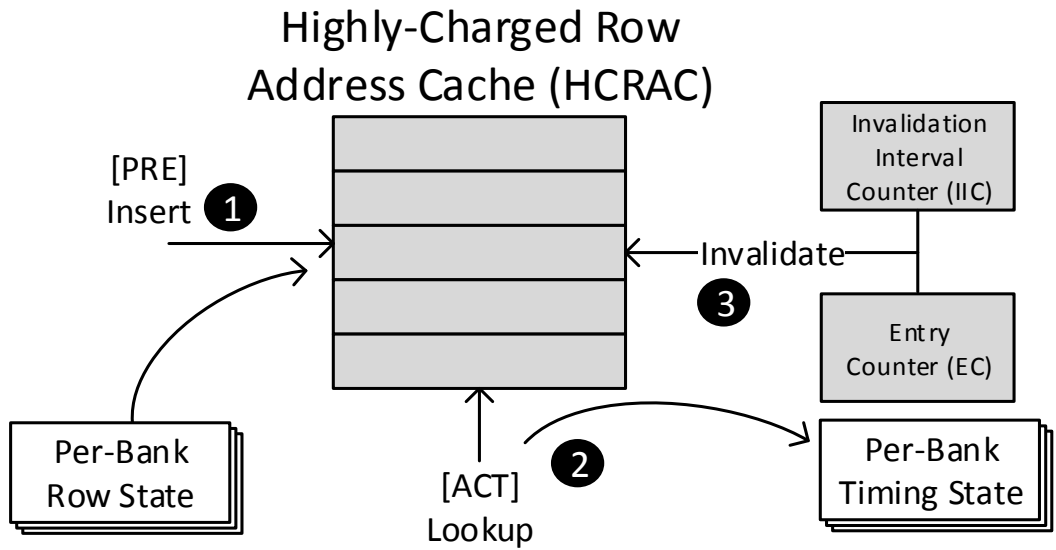}
\caption{Components of the ChargeCache Mechanism} \label{figure:charge_cache}
\end{figure}

\vspace{-3mm}
\subsubsection{Inserting rows into HCRAC}
\label{subsubsection:hcrac_insertion}
When a \textit{PRE} command is issued to a bank, ChargeCache
inserts the address of the row that was activated in the
corresponding bank into the HCRAC \circled{1}. Although the
\textit{PRE} command itself is associated only with the bank
address, the memory controller has to maintain the address of the
row that is activated in each bank (if any row is activated) so
that it can issue appropriate commands when a bank receives a
memory request. ChargeCache obtains the necessary
row address information directly from the memory controller.
Some DRAM
interfaces~\cite{micronDDR3} allow the memory controller to
precharge all banks with a single command. In such cases,
ChargeCache inserts the addresses of the activated rows across
\emph{all} the banks into the HCRAC.

Just like any other cache, HCRAC contains a limited number of
entries. As a result, when a new row address is inserted,
ChargeCache may have to evict an already valid entry from the
HCRAC. While such evictions can potentially result in wasted
opportunity to reduce DRAM latency for some row activations, our
evaluations show that even with a small HCRAC (e.g.,
128-entries), ChargeCache can provide significant performance
improvement (see Section~\ref{section:results}).

\subsubsection{Employing lowered DRAM timing constraints}
\label{subsubsection:employing_lower_params}

To employ lower latency for highly-charged rows, the memory
controller maintains two sets of timing constraints, one for
regular DRAM rows, and another for highly-charged DRAM rows. While we
evaluate the potential reduction in timing constraints that
can be enabled by ChargeCache, we expect the lowered timing
constraints for highly-charged rows to be part of the standard
DRAM specification.

On each \textit{ACT} command, ChargeCache looks up the
corresponding row address in the HCRAC \circled{2}. Upon a
hit, ChargeCache employs lower \textit{tRCD} and \textit{tRAS} for
the subsequent \textit{READ}/\textit{WRITE} and \textit{PRE}
operations, respectively. Upon a miss, ChargeCache employs
the default timing constraints for the subsequent commands.

\subsubsection{Invalidating stale rows from HCRAC}
\label{subsubsection:invalidating_stale}

Unlike conventional caches, where an entry can stay valid as long
as it is not explicitly evicted, entries in HCRAC have to be
invalidated after a specific time interval. This is because as
DRAM cells continuously leak charge, a highly-charged row will no
longer be highly-charged after a specific time interval.

One simple way to invalidate stale entries would be to use a clock
to track time and associate each entry with an expiration time.
Upon a hit in the HCRAC, ChargeCache can check if the entry is
past the expiration time to determine which set of timing
parameters to use for the corresponding row. However, this scheme
increases the storage cost and complexity of implementing
ChargeCache.

We propose a simpler, periodic invalidation scheme that is similar to how
the memory controller issues refresh commands~\cite{liu2012raidr}. Our mechanism
uses
two counters, namely, the \emph{Invalidation Interval Counter}
(IIC) and the \emph{Entry Counter} (EC). We assume that the
HCRAC contains $k$ entries and the number of processor cycles for
which a DRAM row stays highly-charged after a precharge is
$C$. IIC cyclically counts up to $C/k$, and EC cyclically counts
up to $k$. Initially, both IIC and EC are initialized to zero.
IIC is incremented every cycle. Whenever IIC reaches
$C/k$, 1)~the entry in the HCRAC pointed to by EC is
invalidated, 2)~EC is incremented, and 3)~IIC is
cleared. Whenever EC reaches $k$, it is cleared. This
mechanism invalidates every entry in the HCRAC once every $C$
processor cycles. Therefore, it ensures that any valid entry in
the HCRAC indeed corresponds to a highly-charged row. While our
mechanism can prematurely invalidate an entry, our evaluations
show that the loss in performance benefit due to such premature
evictions is negligible.

\subsection{Reduction in DRAM Timing Parameters}
\label{subsection:spice_sim}

We evaluate the potential reduction in \textit{tRCD} and \textit{tRAS} for
ChargeCache using circuit-level SPICE simulations. We implement the DRAM sense
amplifier circuit using $55nm$ DDR3 model
parameters~\cite{rambus} and PTM low-power transistor models~\cite{zhaoptm,
ptmweb}. Figure~\ref{figure:spice} plots the variation in bitline voltage level
during cell activation for different initial charge amounts of the cell.

\begin{figure}
\centering
\includegraphics[width=.95\linewidth]{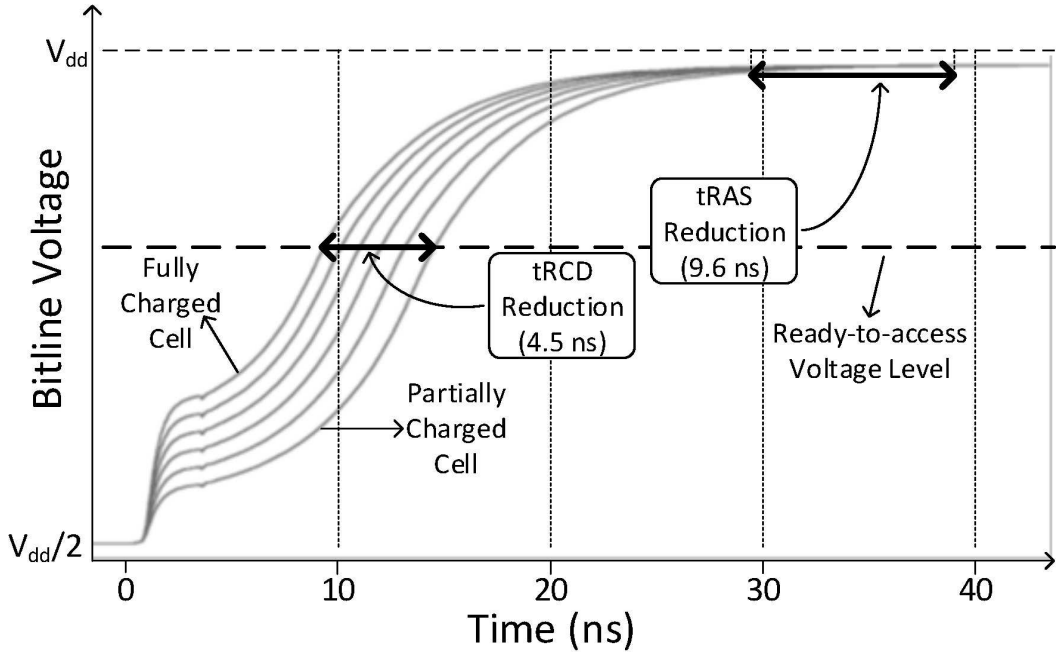}
\caption{Effect of initial cell charge on bitline voltage.}
\label{figure:spice} 
\end{figure}

\setstretch{0.95}

Depending on the initial charge (i.e., voltage level) of the cell, the
bitline voltage increases at different speeds. When the cell is
\emph{fully-charged}, the sense amplifier is able to drive the bitline
voltage to the \emph{ready-to-access voltage level} in only
$10ns$. However, a partially-charged cell (i.e., one
that has not been accessed for $64ms$) brings the bitline voltage up
slower. Specifically, the bitline connected to such a
partially-charged cell reaches the ready-to-access voltage level in
$14.5ns$. Since DRAM timing parameters are dictated
by this worst-case partially-charged state right before the refresh
interval, we can achieve $4.5ns$ reduction in
\textit{tRCD} for a fully-charged cell.  Similarly, the charge of the
cell capacitor is restored at different times depending on the initial
voltage of the cell. For a fully-charged cell, this results in
$9.6ns$ reduction in \textit{tRAS}.

In practice, we expect the DRAM manufacturers to identify the
lowered timing constraints for different caching durations.
Today, DRAM manufacturers test each DRAM chip to determine if it
meets the timing specifications. Similarly, we expect the
manufacturers would also test each chip to determine if it meets
the ChargeCache timing constraints.

\textit{Caching duration} (i.e., how long a row address stays in
ChargeCache) provides a trade-off between ChargeCache hit-rate and the
DRAM access latency reduction. A longer \textit{caching duration}
leads to a longer \textit{Invalidation Interval}. Thus, a row address
stays a longer time in ChargeCache. This creates an opportunity to
increase ChargeCache hit-rate. On the other hand, with a longer
\textit{caching duration}, the amount of charge that remains in DRAM
cells at the end of the duration decreases.  Consequently, the room
for reducing \textit{tRCD} and \textit{tRAS} shrinks. As
Figure~\ref{figure:row_locality} indicates a very high \textit{RLTL}
even with a $0.125ms$ duration, we believe sacrificing
ChargeCache hit-rate for DRAM access latency is a reasonable design
choice. Therefore, we assume a $1ms$ \textit{caching
  duration} and a corresponding 4/8 cycle reduction in
\textit{tRCD}/\textit{tRAS} (determined using SPICE simulations) for a
DRAM bus clocked at 800 MHz frequency. To support our design decision,
we also analyze the effect of various \textit{caching durations} in
Section~\ref{subsubsection:caching_dur}.
\setstretch{1.0}

\section{METHODOLOGY}
\label{section:experimental_setup}

To evaluate the performance of ChargeCache, we use a
cycle-accurate DRAM simulator, Ramulator~\cite{ramulator,
ramulator_web}, in
CPU-trace-driven mode. CPU traces are collected using a
Pintool~\cite{pintool}. Table~\ref{table:system_config} lists the
configuration of the evaluated systems.  We implement the HCRAC
similarly to a 2-way associative cache that uses the LRU policy.

\begin{scriptsize} 
    \begin{table}[h!] 
        \caption{Simulated system configuration}
        \centering 
        \renewcommand{\arraystretch}{1.4} 
        \begin{tabular}{m{3cm} m{9cm}} \hline 
            Processor & 1-8 cores, 4GHz clock frequency, 3-wide issue, 8 MSHRs/core,
                128-entry instruction window\\ \hline 
            Last-level Cache & 64B cache-line, 16-way
                associative, 4MB cache size \\ \hline 
            Memory \hspace{0.5mm} Controller & 64-entry read/write
                request queues, FR-FCFS scheduling policy~\cite{frfcfs,
                zuravleff1997controller}, open/closed row
                policy~\cite{kim2010thread, kim2010} for single/multi core\\ \hline 
            DRAM & DDR3-1600~\cite{micronDDR3}, 800MHz bus frequency, 1/2 channels, 1 rank/channel,
                8 banks/rank, 64K rows/bank, 8KB row-buffer size, tRCD/tRAS  11/28 cycles\\ \hline 
            ChargeCache & 128-entry (672 bytes)/core, 2-way associativity, 
                LRU replacement policy, $1ms$ \textit{caching
                duration}, tRCD/tRAS reduction 4/8 cycles \\ \hline 
        \end{tabular}
        \label{table:system_config} 
    \end{table} 
\end{scriptsize}

For area, power, and energy measurements, we modify
McPAT~\cite{mcpat} to implement ChargeCache using
$22nm$ process technology. We also use
DRAMPower~\cite{drampower} to obtain power/energy results of the
off-chip main memory subsystem. We feed DRAMPower with DRAM
command traces obtained from our simulations using Ramulator.

We run 22 workloads from SPEC CPU2006~\cite{spec2006}, TPC~\cite{tpc}
and STREAM~\cite{stream} benchmark suites. We use
SimPoint~\cite{simpoint} to obtain traces from representative
phases of each application. For single-core evaluations, unless
stated otherwise, we run each workload for 1 billion
instructions. For multi-core evaluations, we use 20 multi-programmed
workloads by assigning a randomly-chosen application to each
core. We evaluate each configuration with its best performing
row-buffer management policy. Specifically, we use the open-row
policy for single-core and closed-row policy for multi-core
configurations. We simulate the benchmarks until each core
executes at least 1 billion instructions. For both single and
multi-core configurations, we first warm up the caches
and ChargeCache by fast-forwarding 200 million cycles.

We measure performance improvement for single-core workloads using
the Intructions per Cycle (IPC) metric. We measure multi-core
performance using the weighted speedup~\cite{snavely2000symbiotic}
metric. Prior work has shown that weighted speedup is a measure of
system throughput~\cite{eyerman2008system}.

\section{EVALUATION}
\label{section:results}

We experimentally evaluate the following mechanisms: 1)
ChargeCache~\cite{hassan2016chargecache}, 2) NUAT~\cite{shin}, which accesses
\emph{only} rows that are \emph{recently-refreshed} at lower latency than the
DRAM standard, 3) ChargeCache + NUAT, which is a combination of ChargeCache and
NUAT~\cite{shin} mechanisms, and 4) Low-Latency DRAM (LL-DRAM)~\cite{lldram},
which is an idealized comparison point where we assume \emph{all rows} in DRAM
can be accessed with low latency, compared to our baseline
DDR3-1600~\cite{micronDDR3} memory, at any time, regardless of when they are
accessed or refreshed.

We primarily use a 128-entry ChargeCache, which provides an effective
trade-off between performance and hardware overhead. We analyze
sensitivity to ChargeCache capacity in
Section~\ref{subsubsection:cc_capacity}. We evaluate LL-DRAM to show
the upper limit of performance improvement that can be achieved by
reducing \textit{tRCD} and \textit{tRAS}. LL-DRAM uses, for all DRAM
accesses, the same reduced values for these timing parameters as we
use for ChargeCache hits. In other words, LL-DRAM is the same as
ChargeCache with a 100\% hit rate.

We compare the performance of our mechanism against the most closely
related previous work, NUAT~\cite{shin}, and also show the benefit of
using both ChargeCache and NUAT together.  The key idea of NUAT is to
access \emph{recently-refreshed} rows at low latency, because these
rows are already highly-charged. Thus, NUAT does not usually access
rows that are recently-\emph{accessed} at low latency, and hence it
does not exploit existing RLTL (Row-Level Temporal Locality) present
in many applications. As we show in Section~\ref{section:motivation},
the fraction of activations that are to rows that are
recently-accessed by the application is much higher than the fraction
of activations that are to rows that are recently-refreshed. In other
words, many workloads have very high RLTL, which is not exploited by
NUAT. As a result, we expect ChargeCache to significantly outperform
NUAT since it can reduce DRAM latency for a much greater fraction of
DRAM accesses than NUAT. To quantitatively prove our expectation that
ChargeCache should widely outperform NUAT, we implement NUAT in
Ramulator using the default 5PB configuration used in~\cite{shin}.

Note that NUAT bins the rows into different latency categories based
on how recently they were refreshed. For instance, NUAT accesses rows
that were refreshed between $0-6ms$ ago with different \textit{tRCD}
and \textit{tRAS} parameters than rows that were refreshed between
$6-16ms$ ago. We determined the different timing parameters of
different NUAT bins using SPICE simulations. Although ChargeCache can
implement a similar approach to NUAT by using multiple {\em caching
  durations}, our \textit{RLTL} results motivate a single
\textit{caching duration} since a row is typically accessed within
$1ms$ (as shown in Section~\ref{section:motivation}). A row that hits
in ChargeCache is always accessed with reduced timings
(Section~\ref{subsection:spice_sim}).

\subsection{Impact on Performance}
\label{subsection:ipc_improvement}

Figure~\ref{figure:ipc} shows the performance of single-core and
eight-core workloads. The figure also includes the number of row
misses per kilo-cycles (RMPKC) to show row activation intensity,
which provides insight into the RLTL of the workload.

\afterpage{
    \clearpage
    \thispagestyle{empty}
    \begin{landscape}
        \centering
        \vspace*{\fill}
        \begin{figure}[htpb]
            \centering
            \begin{subfigure}{\linewidth}
                \centering
                \vspace{-18mm}
                \includegraphics[width=.95\linewidth]{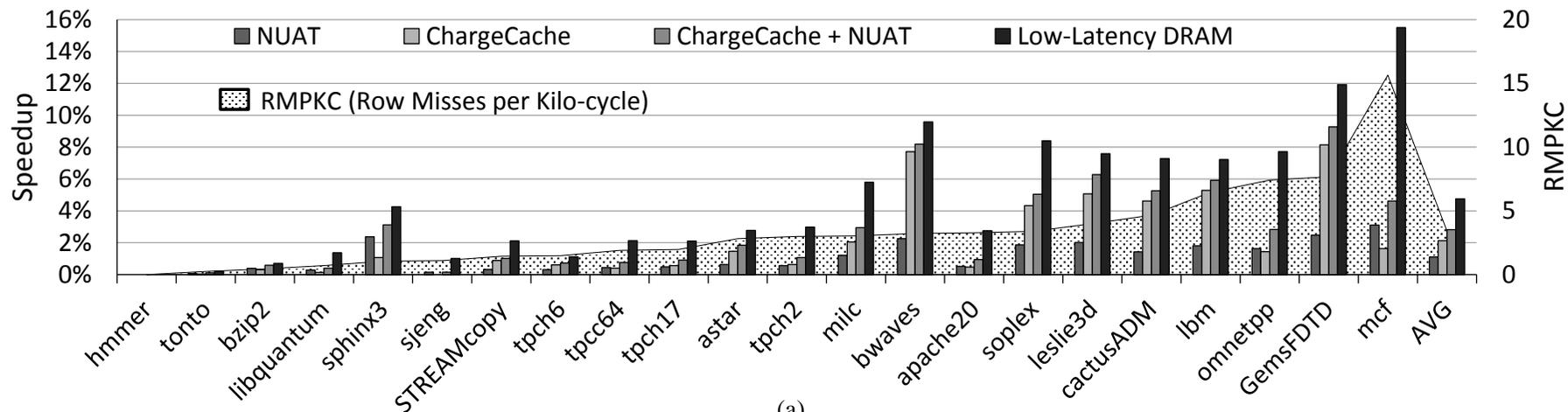}
                \vspace{-5mm}
                \caption{}
                \label{subfigure:ipc_sc}
            \end{subfigure}
            \begin{subfigure}{\linewidth}
                \centering
                \includegraphics[width=.95\linewidth]{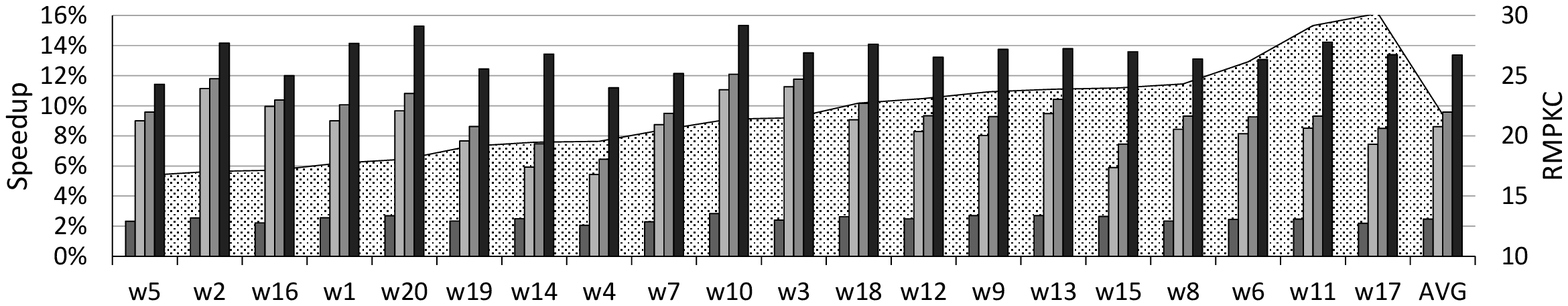}
                \caption{}
                \label{subfigure:ipc_8c}
            \end{subfigure}
            \caption{Speedup with ChargeCache, NUAT and Low-Latency DRAM for
            single-core and eight-core workloads ((a) Single-core workloads, (b)
        Eight-core workloads).}
            \label{figure:ipc}
        \end{figure}
        \null\vfill\centerline{\raisebox{-.6cm}{\thepage}}
    \end{landscape}
    \clearpage
}

\textbf{Single-core.} Figure~\ref{subfigure:ipc_sc}
shows the performance improvement over the baseline system for
single-core workloads. These workloads are sorted in ascending
order of RMPKC. ChargeCache achieves up to 9.3\% (an
average of 2.1\%) speedup.

Our mechanism outperforms NUAT and achieves a speedup close to
LL-DRAM with a few exceptions. Applications that have a wide gap in
performance between ChargeCache and LL-DRAM (such as \emph{mcf,
omnetpp}) access a large number of DRAM rows and exhibit high
row-reuse distance~\cite{kandemir2015}. A high row-reuse distance
indicates that there is large number of accesses to other rows
between two accesses to the same row. Due to this reason, ChargeCache
\emph{cannot} retain the addresses of highly-charged rows until the next
access to that row. Increasing the number of ChargeCache entries
or employing cache management policies aware of reuse
distance or thrashing~\cite{duong2012, seshadri2012evicted,
qureshi2007adaptive} may improve the performance of
ChargeCache for such applications. We leave the evaluation of these
methods for future work. We conclude
that ChargeCache significantly reduces execution time for most
high-RMPKC workloads and outperforms NUAT for all but few
workloads.

\textbf{Eight-core.} Figure~\ref{subfigure:ipc_8c} shows the
speedup on eight-core multiprogrammed workloads. On average,
ChargeCache and NUAT improve performance by 8.6\% and 2.5\%,
respectively. Employing ChargeCache in combination with NUAT
achieves a 9.6\% speedup, which is only 3.8\% less than the
improvement obtained using LL-DRAM. Although the multiprogrammed
workloads are composed of the \emph{same} applications as in
single-core evaluations, we observe much higher performance
improvements among eight-core workloads. The reason is twofold.

First, since multiple cores share a limited capacity LLC,
simultaneously running applications compete for the LLC. Thus,
individual applications access main memory more often, which
leads to higher RMPKC. This makes the workload
performance more sensitive to main memory
latency~\cite{chandra2005predicting, iyer2007qos, kim12}. Second, the
memory controllers receive memory requests from multiple
simultaneously-running applications to a limited number of memory
banks. Such requests are likely to
target different rows since they use separate memory regions and
these regions map to separate rows. Therefore, applications
running concurrently
exacerbate the bank-conflict rate and increase the number of row
activations that hit in ChargeCache.

Overall, ChargeCache improves performance by up to 8.1\% (11.3\%)
and 2.1\% (8.6\%) on average for single-core (eight-core)
workloads. It outperforms NUAT for most of the applications and
using NUAT in combination with ChargeCache improves the
performance slightly further.

\subsection{Impact on DRAM Energy}
\label{subsection:energy_efficiency}

ChargeCache incurs negligible area and power overheads
(Section~\ref{subsection:area_power_overhead}).  Because it
reduces execution time with negligible overhead, it leads to
significant energy savings. Even though ChargeCache increases the
energy efficiency of the entire system, we quantitatively evaluate
the energy savings only for the DRAM subsystem since
Ramulator~\cite{ramulator} does not have a detailed CPU model.

Figure~\ref{figure:energy} shows the average and maximum DRAM
energy savings for single-core and eight-core workloads.
ChargeCache reduces energy consumption by up to 6.9\% (14.1\%)
and on average 1.8\% (7.9\%) for single-core (eight-core)
workloads. We conclude that ChargeCache is effective at improving
the energy efficiency of the DRAM subsystem, as well as the
entire system.

\begin{figure}[!hb]
\centering
\includegraphics[width=.85\linewidth]{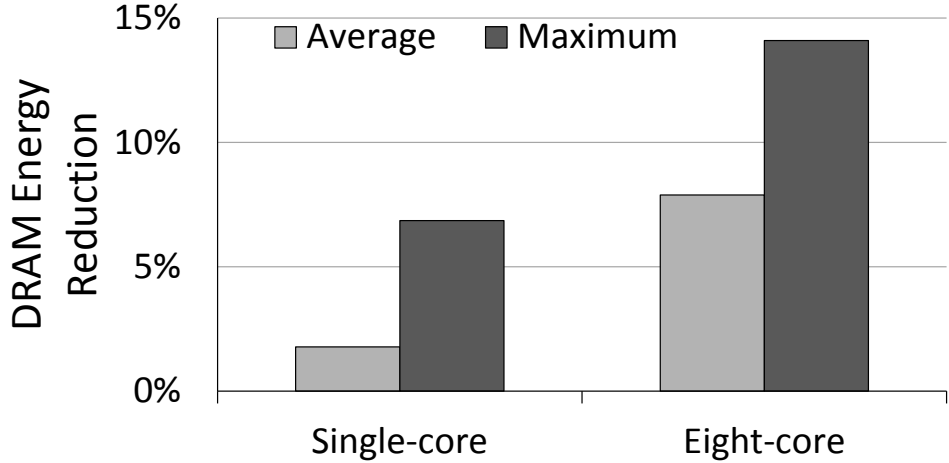}
\caption{DRAM energy reduction of ChargeCache.}
\label{figure:energy}
\end{figure}

\subsection{Area and Power Consumption Overhead}
\label{subsection:area_power_overhead}

HCRAC (Highly-Charged Row Address Cache) is the most area/power
demanding component of ChargeCache. The overhead of \textit{EC}
and \textit{IIC} is negligible since they are just two simple
counters. As we replicate ChargeCache on a per-core and per-memory
channel basis, the total area and power overhead ChargeCache
introduces depends on the number of cores and memory
channels.\footnote{Note that sharing ChargeCache across cores can
  result in even lower overheads. We leave the exploration of such
designs to future work.} The total storage requirement is
given by
Equation~\ref{eq:2}, where \textit{C} are \textit{MC} are the
number of cores and memory channels, respectively.
\textit{LRUbits} depends on ChargeCache associativity. \textit{EntrySize} is
calculated using Equation~\ref{eq:1}, where \textit{R},
\textit{B}, and \textit{Ro} are the number of ranks, banks, and
rows in DRAM, respectively.

\begin{equation}
\label{eq:2}
\resizebox{.8\linewidth}{!}{
$Storage_{bits} = C * MC * Entries * (EntrySize_{bits} + LRU_{bits})$}
\end{equation}

\begin{equation} \label{eq:1}
 \resizebox{.7\linewidth}{!}{$Entry Size_{bits} = log_{2}(R) +
log_{2}(B) +
 log_{2}(Ro) + 1$} 
\end{equation}

\textbf{Area.} Our eight-core configuration has two memory
channels.
This introduces a total of 5376 bytes in storage requirement
for a 128-entry ChargeCache, corresponding to an area of
\unit{0.022}{\milli\meter\squared}. This overhead
is only 0.24\% of the 4MB LLC.

\textbf{Power Consumption.} ChargeCache is accessed on every
\textit{activate} and \textit{precharge} command issued by the memory
controller. On an \emph{activate} command, ChargeCache is searched for
the corresponding row address. On a \emph{precharge} command, the
address of the precharged row is inserted into
ChargeCache. ChargeCache entries are periodically invalidated to
ensure they do not exceed a specified \textit{caching duration}. These
three operations increase dynamic power consumption in the memory
controller, and the ChargeCache storage increases static power
consumption.  Our analysis indicates that ChargeCache consumes
\unit{0.149}{\milli\watt} on average. This is only 0.23\% of the
average power consumption of the entire 4MB LLC. Note that we include
the effect of this additional power consumption in our DRAM energy
evaluations in Section~\ref{subsection:energy_efficiency}. We conclude
that ChargeCache incurs almost negligible chip area and power
consumption overheads.

\subsection{Sensitivity Studies}
\label{subsection:sensitivity_study}

ChargeCache performance depends mainly on two variables:
\textit{HCRAC capacity} and \textit{caching duration}. We observed
that associativity has a negligible effect on ChargeCache
performance. In our experiments, increasing the associativity of
HCRAC from two to full-associativity improved the hit rate by only
2\%. We analyze the hit rate and performance
impact of \emph{capacity} and \textit{caching duration} in more detail.

\subsubsection{ChargeCache capacity}
\label{subsubsection:cc_capacity}

Figure~\ref{figure:cc_hit_rates} shows the average hit rate
versus capacity of ChargeCache for single-core and
eight-core systems. The horizontal dashed lines indicate the maximum
hit rate achievable with an unlimited-capacity ChargeCache.
We observe that 128 entries is a sweet spot between hit rate and
storage overhead.  Such a configuration yields 38\% and 66\%
hit rate for single-core and eight-core systems,
respectively. The storage requirement for a 128-entry ChargeCache
is only 672 bytes per core assuming our two-channel main memory
(see Section~\ref{subsection:area_power_overhead}).

\begin{figure}
\centering
\includegraphics[width=.95\linewidth]{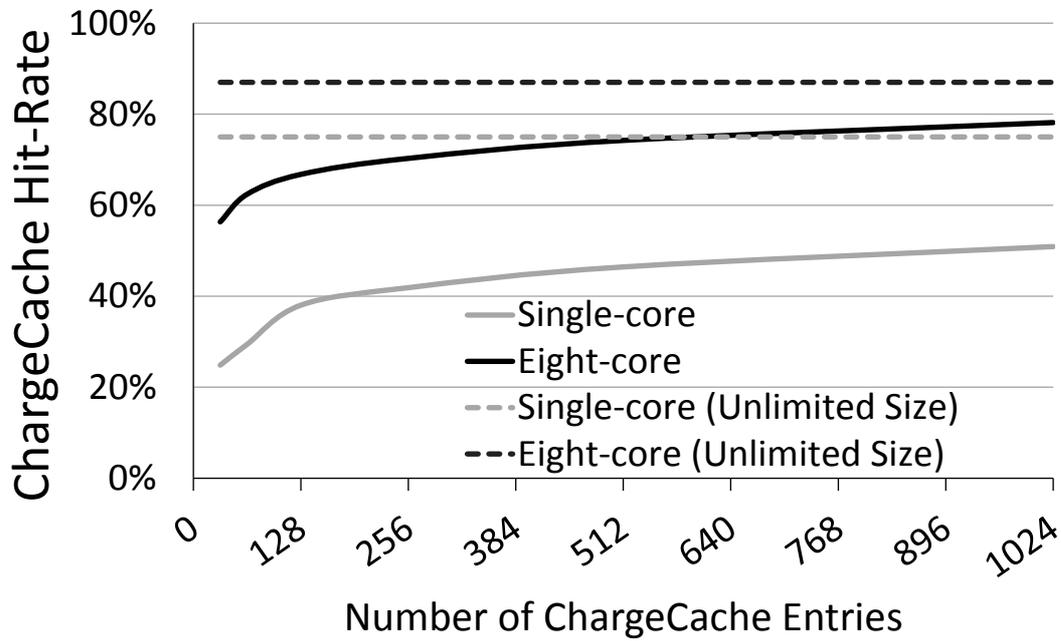}
\caption{ChargeCache hit rate for single-core and eight-core systems at
1ms \textit{caching duration}.}
\label{figure:cc_hit_rates}
\end{figure}

Figure~\ref{figure:ipc_cc_cap} shows the speedup with various
ChargeCache capacities. Larger capacities provide higher
performance thanks to the higher ChargeCache hit rate. However,
they also incur higher hardware overhead.
For a 128-entry capacity (672 bytes per-core), ChargeCache
provides 8.8\% performance improvement, and for a 1024-entry
capacity (5376 bytes per-core) it provides 10.6\% performance
improvement. We conclude that ChargeCache is effective at various
sizes, but its benefits start to diminish at higher capacities.

\begin{figure}[!ht]
\centering
\includegraphics[width=.95\linewidth]{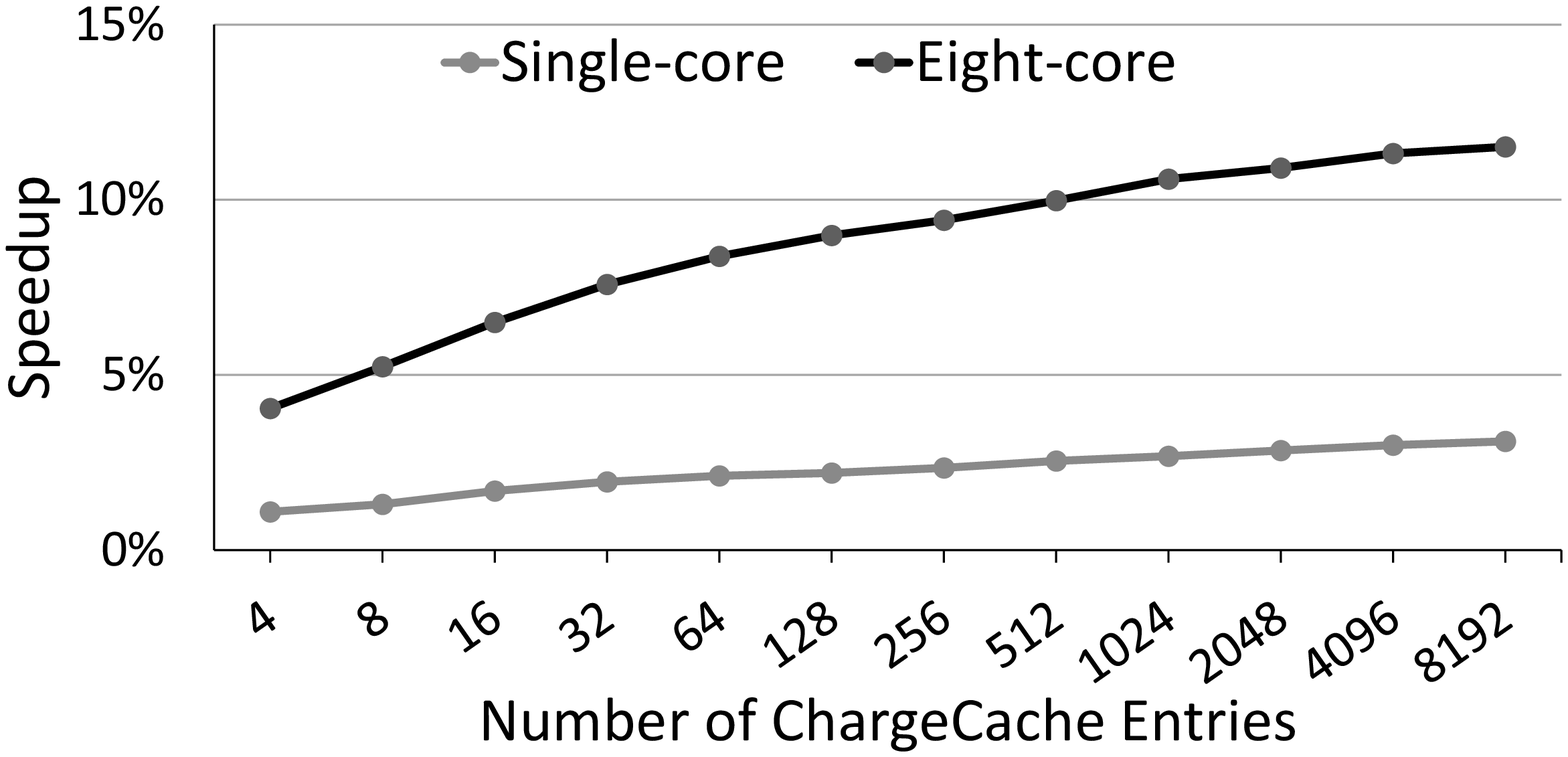}
\caption{Speedup versus ChargeCache capacity.}
\label{figure:ipc_cc_cap}
\end{figure}

\subsubsection{Caching duration} \label{subsubsection:caching_dur}

Increasing the \textit{caching duration} may improve the hit rate by
decreasing the number of invalidated entries. We evaluate several
\textit{caching durations} to determine the
duration value that provides favorable performance. For each
\textit{caching duration}, Table~\ref{table:latency_reduction}
shows the \textit{tRCD} and \textit{tRAS} values which we obtain
from our circuit-level SPICE simulations. We also provide the
default timing parameters used as a baseline in the first row of
the table.

\begin{table}
\caption{tRCD and tRAS for different \textit{caching durations
(determined via SPICE simulations)}}
\centering
\resizebox{.7\linewidth}{!}{
\renewcommand{\arraystretch}{1}
\begin{tabular}{
    m{.28\linewidth} | m{.20\linewidth} | m{.20\linewidth} }
    \textbf{Caching Duration (ms)} & \textbf{tRCD (ns)} & \textbf{tRAS (ns)} \\
\hline N/A (Baseline) & 13.75 & 35\\ \hline 1 & 8 & 22 \\ \hline 4 & 9 & 24 \\
\hline 16 & 11 & 28 \\
  \end{tabular}} \label{table:latency_reduction} \end{table}

Figure~\ref{figure:ipc_caching_dur} shows how ChargeCache speedup and
ChargeCache hit rate vary with different \emph{caching durations}.  We
make two observations. First, increasing the \emph{caching duration}
negatively affects the performance improvement of ChargeCache. This is
because a longer caching duration leads to lower reductions in
\textit{tRCD} and \textit{tRAS} (as
Table~\ref{table:latency_reduction} shows), thereby reducing the
benefit of a ChargeCache hit. Second, ChargeCache hit rate increases
slightly (by about 2\%) for the single-core system but remains almost
constant for the eight-core system when \emph{caching duration}
increases. The latter is due to the large number of bank conflicts in
the 8-core system, as we explained in
Section~\ref{section:motivation}.  With many bank conflicts, the
aggregate number of precharge commands is high and ChargeCache evicts
entries very frequently even with a $1ms$ {\em caching duration}.
Thus, a longer \emph{caching duration} does not have much effect on
hit rate.

\begin{figure}
\centering
\includegraphics[width=.95\linewidth]{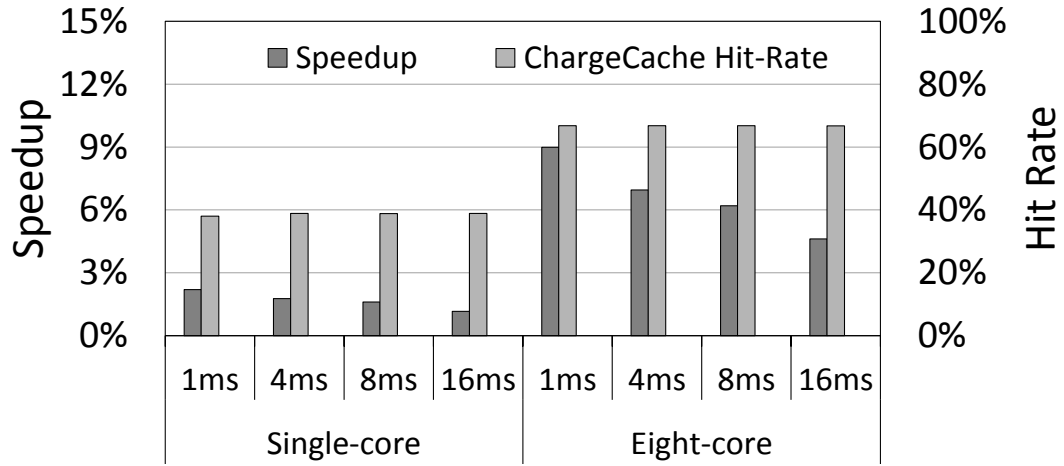}
\caption{Speedup and ChargeCache hit rate for different
\textit{caching durations}}
\label{figure:ipc_caching_dur} 
\end{figure}

We conclude that, with a longer {\em caching duration}, the
improvement in ChargeCache hit rate does not make up for the loss in
the reduction of the timing parameters.  We conclude that ChargeCache
is effective for various \emph{caching durations}, yet the empirically
best \emph{caching duration} is $1ms$, which leads to the highest
performance improvement.

\section{DISCUSSION} 
\label{section:discussion}

\subsection{Temperature Independence}
\label{subsection:temperature_independence}

Charge leakage rate of DRAM cells approximately doubles for every
10$^{\circ}$C increase in the temperature~\cite{lee2015,
khan2014efficacy,mori2005origin, liu2013experimental, qureshi2015avatar}. This
observation can be exploited to lower the DRAM latency when
operating at low temperatures. A previous study, Adaptive Latency
DRAM (AL-DRAM)~\cite{lee2015}, proposes a mechanism to improve
system performance by reducing the DRAM timing parameters at low
operating temperature. It is based on the premise that
DRAM typically does not operate at temperatures close to the
worst-case temperature (85$^{\circ}$~C) even when it is
heavily accessed. However, new 3D-stacked DRAM technologies
such as HMC~\cite{hmc2spec}, HBM~\cite{hbm2013}, WideIO~\cite{dutoit20130} may
operate at significantly higher temperatures due to tight
integration of multiple stack layers~\cite{black,
puttaswamy,lee2016simultaneous}. Therefore, \textit{dynamic
latency scaling} techniques such as AL-DRAM may be less useful in
these scenarios.

ChargeCache is \emph{not} based on the charge difference that occurs due
to temperature dependence. Rather, we exploit the high level of
charge in recently-precharged rows to reduce timing parameters
during later accesses to such rows. After conducting tests to
determine the reduction in timing parameters (for ChargeCache
hits) at \emph{worst-case} temperatures, we find that these
timing parameters can be reduced independently of the operating
temperature. Dynamic latency scaling can still be used in
conjunction with ChargeCache at low temperatures to reduce the
access latency even further.

\subsection{Applicability to Other DRAM Standards} \label{subsection:other_dram}

Although we evaluate only DDR3-based main memory within the scope of
this thesis, implementing ChargeCache for other DRAM standards is
straightforward. In theory, ChargeCache is applicable to any memory
technology where cells are volatile (leak charge over time). However,
the memory interface can prevent the implementation of ChargeCache
entirely in the memory controller. For example, RL-DRAM~\cite{rldram}
is a DRAM type incompatible with ChargeCache. In RL-DRAM, read and
write operations are directly handled by \textit{READ} and
\textit{WRITE} commands without explicitly activating and precharging
DRAM rows.  Hence, the RL-DRAM memory controller does not have control
over the activation delay of the rows and the timing parameters
\textit{tRCD} and \textit{tRAS} do not exist.

However, ChargeCache can be used with to a large set of
specifications derived from DDR (DDRx, GDDRx, LPDDRx, etc.) in a
manner similar to the mechanism described in this work, without
modifying the DRAM architecture at all. All of these memories
require \textit{ACT} and \textit{PRE} commands to explicitly open
and close DRAM rows.  Using ChargeCache with 3D-stacked
memories~\cite{loh, lee2016simultaneous} such as
WideIO~\cite{dutoit20130}, HBM~\cite{hbm2013} and
HMC~\cite{hmc2spec} is also straightforward. The difference is
that the DRAM controller, and hence ChargeCache, may be
implemented in the logic layer of the 3D-stacked memory chip
instead of the processor chip.

\section{RELATED WORK}
\label{section:related_work}

To our knowledge, this work is the first to (\textit{i}) show
that applications typically exhibit significant \textit{Row-level
Temporal Locality (RLTL)} and (\textit{ii}) exploit this locality
to improve system performance by reducing the latency of
requests to recently-accessed rows.

We have already qualitatively and quantitatively (in
Sections~\ref{section:motivation} and~\ref{section:results}) compared
ChargeCache to NUAT~\cite{shin}, which reduces access latency to
\emph{only} recently-refreshed rows. We have shown that ChargeCache can
provide significantly higher average latency reduction than NUAT
because RLTL is usually high, whereas the fraction of accesses to
rows that are recently-refreshed is typically low.

Other previous works have proposed  techniques to reduce
performance degradation caused by long DRAM latencies. They
focused on 1) enhancing the DRAM, 2) exploiting variations in
manufacturing process and operating conditions, 3) developing
several memory scheduling policies. We briefly summarize how
ChargeCache differs from these works.

\textbf{Enhancing DRAM Architecture.} Lee at al.  propose
Tiered-Latency DRAM (TL-DRAM)~\cite{lee} which divides each subarray
into near and far segments using isolation transistors. With TL-DRAM,
the memory controller accesses the near segment with lower latency
since the isolation transistor reduces bitline capacitance in that
segment. Our mechanism could be implemented on top of TL-DRAM to
reduce the access latency for both the near and far segment. Kim et
al. unlock parallelism among subarrays at low cost with
SALP~\cite{kim12}. The goal of SALP is to reduce DRAM latency by
providing more parallelism to reduce the impact of bank conflicts. O
et al~\cite{son2014} propose a DRAM architecture where sense
amplifiers are decoupled from bitlines to mitigate precharge
latency. Choi et al~\cite{choi} propose to utilize multiple DRAM cells
to store a single bit when sufficient DRAM capacity is available. By
using multiple cells, they reduce activation, precharge and refresh
latencies. Other works~\cite{gulur2012, son, zhang2014, seshadri2013,
  seshadri2015fast, seshadri2015gather, chang2014, chang2016} also
propose new DRAM architectures to lower DRAM latency.

Unlike ChargeCache, all these works require changes to the DRAM
architecture itself. The approaches taken by these works are
largely orthogonal and ChargeCache could be implemented together
with any of these mechanisms to further improve the DRAM latency.

\textbf{Exploiting Process and Operating Condition Variations.}
Recent studies~\cite{lee2015,chandrasekar2014} proposed methods
to reduce the safety margins of the DRAM timing parameters when
operating conditions are appropriate (i.e., not worst-case).
Unlike these works, ChargeCache is largely independent of
operating conditions like temperature, as discussed in
Section~\ref{subsection:temperature_independence}, and is
orthogonal to these latency reduction mechanisms. 

\textbf{Memory Request Scheduling Policies.} Memory request
scheduling policies (e.g.,~\cite{lee2010dram, frfcfs, mutlu2007stall,
mutlu08, kim2010, kim2010thread, subramanian2014blacklisting,
subramanian2015application, subramanian2013mise, usui2016dash})
reduce the average DRAM access latency by improving DRAM
parallelism, row-buffer locality and fairness in especially
multi-core systems. 
ChargeCache can be employed in conjunction with the scheduling
policy that best suits the application and the underlying
architecture.

\section{CONCLUSION}
\label{section:conclusion}

We introduce ChargeCache, a new, low-overhead mechanism that
dynamically reduces the DRAM timing parameters for
recently-accessed DRAM rows. ChargeCache exploits two key
observations that we demonstrate in this work: 1) a
recently-accessed DRAM row has cells with high amount of charge
and thus can be accessed faster, 2) many applications repeatedly access
rows that are recently-accessed.

Our extensive evaluations of ChargeCache on both single-core and
multi-core systems show that it provides significant performance
benefit and DRAM energy reduction at very
modest hardware overhead. ChargeCache requires no modifications
to the existing DRAM chips and occupies only a small area on the
memory controller.

We conclude that ChargeCache is a simple yet efficient mechanism to dynamically
reduce DRAM latency, which significantly improves both the performance and
energy
efficiency of modern systems.

\vfill
\newpage






\bibliographystyle{acm}

\addcontentsline{toc}{section}{\textbf{REFERENCES}}

\bibliography{Etuthesis}

\vfill
\newpage

}

\end{document}